\documentclass[11pt]{article}
\usepackage[margin=1in]{geometry}
\usepackage{amsmath,amssymb,amsthm}
\usepackage{enumitem}
\usepackage{cite}
\usepackage{xcolor}
\usepackage{hyperref}

\newcommand{\Asf}{\mathsf{A}}
\newcommand{\Ksf}{\mathsf{K}}
\newcommand{\Nsf}{\mathsf{N}}
\newcommand{\Msf}{\mathsf{M}}
\newcommand{\Rsf}{\mathsf{R}}
\newcommand{\Bsf}{\mathsf{B}}
\newcommand{\qsf}{\mathsf{q}}
\newcommand{\rsf}{\mathsf{r}}
\newcommand{\FF}{\mathbb{F}}
\newcommand{\Ic}{\mathcal{I}}
\newcommand{\Wc}{\mathcal{W}}

\newcommand{\Sc}{\mathcal{S}}
\newcommand{\Gc}{\mathcal{G}}
\newcommand{\Tc}{\mathcal{T}}
\newcommand{\dv}{\mathbf{d}}
\newcommand{\Gm}{\mathbf{G}}
\newcommand{\Wm}{\mathbf{W}}

\newcommand{\cv}{\mathbf{c}}
\newcommand{\fv}{\mathbf{f}}
\newcommand{\ev}{\mathbf{e}}
\newcommand{\gv}{\mathbf{g}}
\newcommand{\hv}{\mathbf{h}}
\newcommand{\pv}{\mathbf{p}}
\newcommand{\uv}{\mathbf{u}}
\newcommand{\xv}{\mathbf{x}}
\newcommand{\Hom}{\mathsf{P}}
\newcommand{\Sym}{\mathsf{S}}

\newtheorem{thm}{Theorem}
\newtheorem{cor}{Corollary}
\newtheorem{lem}{Lemma}
\newtheorem{prop}{Proposition}
\theoremstyle{definition}
\newtheorem{defn}{Definition}
\theoremstyle{remark}
\newtheorem{rem}{Remark}

\title{Hotplug Coded Caching: Embedding the YMA and Interference Elimination Schemes via Reed--Muller Codes}
\author{Daniela Tuninetti\thanks{University of Illinois Chicago, Chicago, IL 60607, USA; Email: danielat@uic.edu. This work was funded in part by NSF under award 2312229.}}
\date{}

\begin{document}
\maketitle

\begin{abstract}
In the hotplug coded caching model, $K$ users fill their caches of size $M$ as a function of $N$ files, but only an unknown subset of $A<K$ users is active in the delivery phase. We recently showed that (i) under uncoded placement, offline users cause a strict load penalty at intermediate memory sizes, so that coded placement is needed; and (ii) for $AM\leq N$, every classical scheme with row-wise placement, including the scheme of Gomez-Vilardebo (GV), can be embedded in a Reed-Solomon code, which removes the penalty. In this note, we show that every point of the scheme of Yu, Maddah-Ali and Avestimehr (YMA) for a classical system with $A$ users is achievable in the hotplug model for every $K\geq A$. Each user is associated with a hyperplane and each file with a homogeneous polynomial; each user caches the restriction of the polynomial to its hyperplane, and the subfiles seen by the active users are the values of the polynomial at the points where their hyperplanes meet. This is a projective Reed-Muller code, which reduces to the Reed-Solomon code when each subfile is cached by one user. We then show, by an existence argument, that every classical scheme whose cache is a linear compression of the Maddah-Ali-Niesen cache and whose correctness holds for generic coefficients can also be embedded over a sufficiently large field; this includes the interference elimination (IE) schemes of Tian and Chen. These embeddings complement the Reed-Solomon one, which allows any row-wise placement and any delivery. Together with past work, our results show that every point attained by a GV, YMA or IE scheme ($N\leq A$ for IE), and every corner point of the linear tradeoff for $A=N=3$, is achievable in the hotplug model with a load that does not depend on $K$. Whether the optimal hotplug tradeoff depends only on $A$ remains open, but these results support a positive answer.
\end{abstract}

\section{Introduction}
Coded caching~\cite{MAN} studies a server with $\Nsf$ files connected through a shared link to $\Ksf$ users with caches of size $\Msf$ files. We call a system \emph{classical} if all the users that take part in the placement phase also take part in the delivery phase, and \emph{hotplug} if only $\Asf$ of the $\Ksf>\Asf$ users that took part in the placement phase are active in the delivery phase, where the set of active users is unknown at the time of placement.

The hotplug model was introduced by Ma and Tuninetti~\cite{MT2022hotplug}, who proposed schemes with a placement phase based on Maximum Distance Separable (MDS) codes. In all regimes where these schemes could be shown to be optimal, the optimal tradeoff of the hotplug model coincides with that of a classical system with $\Asf$ users; in particular, it does not depend on $\Ksf$. Whether this holds in general was posed as an open question~\cite{MT2022hotplug,MT2024arxiv}, and it remains open.

To make progress on this open question, in~\cite{DT2027ICCsubmission} we showed that, under uncoded placement, this is not the case: the optimal scheme under uncoded placement is the natural extension of the scheme of Yu, Maddah-Ali and Avestimehr (YMA)~\cite{YMA}, in which the offline users are assigned the demands of active users, and its load is strictly larger than that of a classical system with $\Asf$ users at intermediate memory sizes. Hence, to attain the classical load in the hotplug model, coded placement is needed in general.
The coded placement schemes in~\cite{DT2027ICCsubmission} cover the small memory regime $\Asf\Msf\leq\Nsf$, where a classical system with $\Asf$ users admits \emph{row-wise placements} (RWP), meaning that each file is split into $\Asf$ rows, and the cache of user $a\in[\Asf]$ is a function of row $a$ of every file. Any such classical RWP scheme can be embedded in a Reed--Solomon code: in the hotplug model, user $k\in[\Ksf]$ caches the same function of a coded row, namely the evaluation at a point $\alpha_k$ of the polynomial whose coefficients are the rows; once the active users are known, their coded rows are the rows of an invertible transformation of the library, and the classical delivery applies~\cite[Theorem~2]{DT2027ICCsubmission}. Thus, both the scheme of G\'omez-Vilardeb\'o (GV)~\cite{gomez2018fundamental} and the YMA point with $\Asf\Msf=\Nsf$ (i.e., one cached row per file) are achievable in the hotplug model for every $\Ksf\geq\Asf$. This note shows that something similar can be done at larger memory sizes.

\subsection{Contributions}
We give three results:
\begin{enumerate}
\item
First, an explicit construction: every YMA point of a classical system with $\Asf$ users can be embedded in the hotplug model by means of a Reed--Muller code (Theorem~\ref{thm:RM}); hence, the YMA tradeoff for $\Asf$ users is achievable in the hotplug model for every $\Ksf\geq\Asf$. 
\item
Second, an existence result: on top of the same construction, a generic linear compression of the MAN cached data embeds every classical scheme whose cache is a linear compression of a MAN cache and whose correctness holds for generic coefficients (Theorem~\ref{thm:generic}); this includes the IE scheme~\cite{tian2018caching}. 
As a consequence, the hotplug tradeoff with $4$ active users and $2$ files equals the classical one wherever the latter is known (Corollary~\ref{cor:42}).
\item
Finally, for $\Asf=\Nsf=3$, we give a placement, row-wise up to one common subfile per file, that attains the point $(\frac35,\frac32)$ of Cao and Xu~\cite{caoxu2020} and embeds via the Reed--Solomon code (Proposition~\ref{prop:N=A=3,M=3/5}); to the best of our knowledge, no scheme that attains this point is known for other values of $\Nsf$ and $\Asf$. As a consequence, under linear coding in both placement and delivery, the hotplug tradeoff with $3$ active users and $3$ files equals the classical one for every $\Ksf\geq3$ (Proposition~\ref{prop:N=A=3,M=3/5}).
\end{enumerate}

\subsection{Paper organization}
We first state the hotplug model and existing results (Section~\ref{sec:known}) and summarize novel results (Section~\ref{sec:main}), then explain the explicit construction by examples (Sections~\ref{sec:example A=N=3} and~\ref{sec:example N=3 A=4}). We then prove Theorem~\ref{thm:RM} (Section~\ref{sec:proof}) for projective Reed--Muller codes, and also give an equivalent description in terms of symmetric polynomials. We then explain why an exact embedding is limiting and prove Theorem~\ref{thm:generic} (Section~\ref{sec:generic}). Finally, we settle the linear-optimal point $(\Msf,\Rsf)=(\frac35,\frac32)$ for $\Asf=\Nsf=3$ (Section~\ref{sec:open}).

\subsection{Notation}
Calligraphic symbols denote sets, bold symbols vectors and matrices, and sans-serif symbols system parameters.
$\binom{a}{b}=0$ unless $a\geq b\geq0$.
For integers $a\leq b$, $[a:b]:=\{a,\ldots,b\}$ and $[b]:=[1:b]$.
For a set $\Gc$, $\Omega_\Gc^t:=\{\Tc\subseteq\Gc: |\Tc|=t\}$.
For a set $\Ic\subseteq\mathbb{N}$, $X_\Ic:=(X_n : n\in \Ic)$,
and $\dv[\Ic]$ is the sub-vector of $\dv$ restricted by $\Ic$.

\section{Model and Known Results}
\label{sec:known}

\subsection{Hotplug model}
A server has $\Nsf$ independent files $F_1,\ldots,F_\Nsf$, each of $\Bsf$ i.i.d.\ uniform symbols over a finite field $\FF_\qsf$; for entropies, logarithms are in base $\qsf$. In the placement phase, user $k\in[\Ksf]$ stores $Z_k$, a function of the files, with $H(Z_k|F_{[\Nsf]})=0$ and $H(Z_k)\leq\Msf\Bsf$. In the delivery phase, the users in $\Ic\in\Omega^\Asf_{[\Ksf]}$ are active, and active user $k$ requests file $d_k\in[\Nsf]$; the server sends $X$, with $H(X|F_{[\Nsf]},\Ic,\dv[\Ic])=0$ and $H(X)\leq\Rsf\Bsf$, and every active user decodes its requested file as $H(F_{d_k}|X,Z_k,\Ic,\dv[\Ic])=0$ for all $k\in\Ic$. The pair $(\Msf,\Rsf)$ is achievable if there is a scheme, for some $\Bsf$ and $\qsf$, that satisfies these constraints for every $\Ic\in\Omega^\Asf_{[\Ksf]}$ and every $\dv[\Ic]\in[\Nsf]^\Asf$. We denote by $\Rsf^\star_{(\Asf,\Ksf,\Nsf)}(\Msf)$ the optimal hotplug worst-case load, i.e., the infimum of $\Rsf$ over achievable pairs $(\Msf,\Rsf)$. The case $\Ksf=\Asf$ is the classical coded caching model.

A placement is \emph{uncoded} if each user stores a subset of the symbols of the files, without coding. We denote by $\Rsf^\star_{{\rm u}(\Asf,\Ksf,\Nsf)}(\Msf)$ the optimal hotplug worst-case load under uncoded placement.

A scheme is \emph{linear} if every cache $Z_k$, $k\in[\Ksf]$, and every transmission $X$, for every $\Ic$ and $\dv[\Ic]$, is a linear function of $(F_1,\ldots,F_\Nsf)$ over $\FF_\qsf$, i.e., placement and delivery are both linear codes. We denote by $\Rsf^\star_{{\rm lc}(\Asf,\Ksf,\Nsf)}(\Msf)$ the optimal hotplug worst-case load over linear schemes, over all $\Bsf$ and $\qsf$.

Since uncoded placement is linear, and the optimal delivery under uncoded placement is linear (Theorem~\ref{thm:uncoded}), we have $\Rsf^\star_{(\Asf,\Ksf,\Nsf)}\leq\Rsf^\star_{{\rm lc}(\Asf,\Ksf,\Nsf)}\leq\Rsf^\star_{{\rm u}(\Asf,\Ksf,\Nsf)} \leq \rsf:=\min(\Asf,\Nsf)$.

\subsection{Known Results for classical coded caching}

\subsubsection{The YMA scheme for classical coded caching}
For a classical system with $\Asf$ users and $t\in[0:\Asf]$, the placement of Maddah-Ali and Niesen (MAN)~\cite{MAN} splits each file into $\binom{\Asf}{t}$ subfiles, $F_n=(F_{n,\Wc}:\Wc\in\Omega^t_{[\Asf]})$, and user $a$ stores $F_{n,\Wc}$ for all $n$ and all $\Wc\ni a$. With the delivery of~\cite{YMA}, this achieves
\begin{align}
(\Msf^{\rm YMA}_t,\Rsf^{\rm YMA}_t)=\Big(\frac{\Nsf t}{\Asf},\ \frac{\binom{\Asf}{t+1}-\binom{\Asf-\rsf}{t+1}}{\binom{\Asf}{t}}\Big),\quad t\in[0:\Asf],\quad \rsf:=\min(\Asf,\Nsf),
\label{eq:YMA}
\end{align}
and the lower convex envelope of these points is optimal under uncoded placement~\cite{YMA,WTP}.

\subsubsection{The GV scheme for classical coded caching}
The GV scheme~\cite{gomez2018fundamental} for $\Asf$ users has a RWP: it further splits each row into pieces indexed by the $g$-subsets of $[\Nsf]$ and caches sums of pieces of $g$ distinct files of the same row; it achieves the lower convex envelope of $(0,\rsf)$ and the points
\begin{align}
(\Msf^{\rm GV}_g,\Rsf^{\rm GV}_g) =\Big(\frac{\Nsf}{\Asf g},\ \rsf-\frac{\Nsf(\rsf+1)}{\Asf(g+1)}\frac{\binom{\rsf}{g}}{\binom{\Nsf}{g}}\Big),\quad g\in[\Nsf],\quad \rsf:=\min(\Asf,\Nsf).
\label{eq:GV}
\end{align}
For $g=\Nsf\leq\Asf$, this is the CFL point $(\frac1\Asf,\Nsf(1-\frac1\Asf))$~\cite{chen2016fundamental}, which caches $\sum_nF_{n,k}$.
For $g=1$, this is the YMA point with $t=1$, $(\frac\Nsf\Asf,\rsf-\frac{\rsf(\rsf+1)}{2\Asf})$, which caches the row itself.
The line through $(0,\Nsf)$ and the CFL point is optimal for $\Nsf\leq\Asf$~\cite{chen2016fundamental}.
The line through the CFL point and the GV point with $g=\Nsf-1$, i.e., $\Msf=\frac{\Nsf}{\Asf(\Nsf-1)}$, is optimal for $\lceil(\Asf+1)/2\rceil\leq\Nsf\leq\Asf$~\cite{kumar2021exact}; for $\Nsf=\Asf$, the optimality of that GV point was shown in~\cite{gomez2018fundamental}.

\subsubsection{The IE scheme for classical coded caching}
For a classical system with $\Asf$ users, $\Nsf\leq\Asf$ and $t\in[0:\Asf]$, the IE scheme uses the same split as the MAN placement, $F_n=(F_{n,\Wc}:\Wc\in\Omega^t_{[\Asf]})$, but user $a$ does not store its $P:=\Nsf\binom{\Asf-1}{t-1}$ subfiles $(F_{n,\Wc}:n\in[\Nsf],\ \Wc\ni a)$; it stores fewer linear combinations of them, namely the parity symbols of a systematic code (a rank-metric code in~\cite{tian2018caching}) that encodes them. These combinations mix subfiles of different files and with different indices $\Wc$. In the delivery, the server sends uncoded the subfiles that are not present at any user requesting the same file, together with parity symbols of MDS codes that encode groups of subfiles of the same file; each user first uses the received symbols to eliminate, from its cached combinations, the subfiles of the files it does not request (the interference), and then decodes the subfiles of its requested file. 
This achieves
\begin{align}
(\Msf^{\rm IE}_t,\Rsf^{\rm IE}_t)=\Big(\frac{t\,[(\Nsf-1)t+\Asf-\Nsf]}{\Asf(\Asf-1)},\ \frac{\Nsf(\Asf-t)}{\Asf}\Big),\quad t\in[0:\Asf].
\label{eq:IE}
\end{align}
Correctness reduces to the full rank of certain matrices, whose determinants are polynomials in the coefficients of the cached combinations; these polynomials are not identically zero, and generic coefficients over a sufficiently large field can replace the rank-metric code, with the same MDS codes in the delivery~\cite[Section~VI-D]{tian2018caching}.

IE improves on YMA at small memory sizes; for $t=1$ it is the point of Chen, Fan and Letaief (CFL)~\cite{chen2016fundamental}, and for $(\Asf,\Nsf)=(4,2)$ and $t=2$ it is $(\frac23,1)$, which is optimal~\cite{tian2018symmetry}. 

\subsection{Known Results for hotplug coded caching}
Our past work showed that the following points of a classical system with $\Asf$ users are achievable in the hotplug model for every $\Ksf\geq\Asf$: the CFL point for $\Asf\geq\Nsf$~\cite[Scheme~new2]{MT2022hotplug} (HT2 in~\cite{MT2024arxiv}), the YMA point with $t=1$~\cite[Scheme~new1]{MT2022hotplug} (HT1 with $t=1$ in~\cite{MT2024arxiv}), and the YMA point with $t=\Asf-1$ for $\Asf\geq3$~\cite[Scheme~HT3]{MT2024arxiv}.  In~\cite[Scheme~new1]{MT2022hotplug} we also added an MDS layer to the MAN placement with $t\geq2$, which attains the YMA load of a classical system with $\Asf$ users, but with a memory size that depends on both $\Asf$ and $\Ksf$.

In particular, the YMA+ scheme~\cite{MT2022hotplug} uses the MAN placement for $\Ksf$ users with parameter $t\in[0:\Ksf]$ and the YMA delivery for $\Ksf$ users, in which each offline user is assigned the demand of an active user; it achieves
\begin{align}
(\Msf^{\rm YMA+}_t,\Rsf^{\rm YMA+}_t)=\Big(\frac{\Nsf t}{\Ksf},\ \frac{\binom{\Ksf}{t+1}-\binom{\Ksf-\rsf}{t+1}}{\binom{\Ksf}{t}}\Big),\quad t\in[0:\Ksf],\quad \rsf:=\min(\Asf,\Nsf),
\label{eq:YMA+}
\end{align}
and we denote by $\Rsf^{(\rm YMA+)}_{(\Asf,\Ksf,\Nsf)}(\Msf)$ the lower convex envelope of these points. 

\begin{thm}[Optimality of YMA+ to within a factor of two~\cite{MT2022hotplug,MT2024arxiv}]
For an $(\Asf,\Ksf,\Nsf)$ hotplug model, 
\begin{align}
\frac{\Rsf^{(\rm dec)}_{(\Asf,\Nsf)}}{2.00884}
\stackrel{\rm(i)}{\leq}
    \Rsf^\star_{(\Asf,\Asf,\Nsf)}
\leq\Rsf^\star_{(\Asf,\Ksf,\Nsf)}
\leq\Rsf^{(\rm YMA+)}_{(\Asf,\Ksf,\Nsf)}
\stackrel{\rm(ii)}{\leq}
     \Rsf^{(\rm dec)}_{(\Asf,\Nsf)},
\label{eq:bounds}
\end{align}
where the lower bounds depend only on $\Asf$, the inequalities marked as (i) and (ii) are from~\cite{YMAfactor2}, and
\begin{align}
\Rsf^{(\rm dec)}_{(\Asf,\Nsf)} = \lim_{\Ksf\to\infty}\Rsf^{(\rm YMA+)}_{(\Asf,\Ksf,\Nsf)}
= \frac{1-\mu}{\mu}\big(1-(1-\mu)^\rsf\big), \quad \mu = \Msf/\Nsf,
\end{align}
is the load with decentralized placement.
The chain of inequalities in~\eqref{eq:bounds} shows that YMA+ is optimal to within a factor of two~\cite{MT2024arxiv}.
\end{thm}

\begin{thm}[Optimality of YMA+ under uncoded placement~\cite{DT2027ICCsubmission}]
\label{thm:uncoded}
For an $(\Asf,\Ksf,\Nsf)$ hotplug model, $\Rsf^\star_{{\rm u}(\Asf,\Ksf,\Nsf)}(\Msf)=\Rsf^{(\rm YMA+)}_{(\Asf,\Ksf,\Nsf)}(\Msf)$ for all $\Msf\in[0,\Nsf]$.
\end{thm}

\begin{thm}[Reed--Solomon embedding of row-wise placement schemes~\cite{DT2027ICCsubmission}]
\label{thm:RS}
Let a classical scheme for $\Asf$ users with RWP, over a field $\FF_{\qsf_0}$, achieve $(\Msf,\Rsf)$, with possibly nonlinear cache placement and delivery functions. Then $(\Msf,\Rsf)$ is also achievable in the $(\Asf,\Ksf,\Nsf)$ hotplug model for every $\Ksf\geq\Asf$, over any extension field $\FF_\qsf$ of $\FF_{\qsf_0}$ with $\qsf\geq\Ksf-1$.
\end{thm}

\begin{cor}[Application of Theorem~\ref{thm:RS}: GV in the hotplug model]
\label{cor:GV}
For every $\Ksf\geq\Asf$, the lower convex envelope of the GV points~\eqref{eq:GV} of a classical system with $\Asf$ users is achievable in the hotplug model with $\Asf$ active users out of $\Ksf$.
This also includes YMA for $t=1$, which is the GV point for $g=1$.
\end{cor}

\section{Main Results}
\label{sec:main}

\subsection{Part~1: exact embedding}
We consider the following class of placements, which contains the MAN placement and the GV placement.
\begin{defn}[Linear $t$-subset placement]
\label{def:tsubset}
For a classical system with $\Asf$ users, fix $t\in[\Asf]$.
Each file is split into $\binom{\Asf}{t}$ subfiles of equal size as $F_n=(F_{n,\Wc}:\Wc\in\Omega^t_{[\Asf]})$ with $F_{n,\Wc}\in\FF_\qsf^{1\times\Bsf/\binom{\Asf}{t}}$; define $\fv_{\Wc}:=[F_{1,\Wc},\ldots,F_{\Nsf,\Wc}]\in\FF_\qsf^{1\times\Nsf\Bsf/\binom{\Asf}{t}}$ as the concatenation of the subfiles indexed by $\Wc$ of all files.
A placement is a linear $t$-subset placement if
\begin{align}
Z_a=\big(\,\mathcal{L}[\fv_{\Wc}]:\ \Wc\in\Omega^t_{[\Asf]},\ \Wc\ni a\big),\quad a\in[\Asf],
\label{eq:tsubset}
\end{align}
where $\mathcal{L}:\FF_\qsf^{1\times\Nsf\Bsf/\binom{\Asf}{t}}\to\FF_\qsf^{1\times\Msf\Bsf/\binom{\Asf-1}{t-1}}$ is a linear operator, the same for every $\Wc$ and every user; hence the cache size is $\Msf\Bsf$.
\end{defn}

Definition~\ref{def:tsubset} does not allow combinations of subfiles with different indices $\Wc$; these are considered in Definition~\ref{def:compression}.
The MAN placement is the case where $\mathcal{L}$ is the identity, i.e., $\Msf=\Nsf t/\Asf$.
The CFL placement~\cite{chen2016fundamental}, in which user $a$ stores $\sum_nF_{n,\{a\}}$, is the case $t=1$ and $\mathcal{L}[x_1,\ldots,x_\Nsf]=x_1+\cdots+x_\Nsf$, i.e., $\Msf=1/\Asf$.

Note: the placement of Proposition~\ref{prop:N=A=3,M=3/5} in Section~\ref{sec:open} for $\Asf=\Nsf=3$ and $\Msf=\frac35$ is not of the form of Definition~\ref{def:tsubset}, because each cached symbol combines private subfiles of one user ($t=1$) with a subfile common to all users ($t=3$). 

\begin{thm}[Reed--Muller embedding]
\label{thm:RM}
Let a classical scheme for $\Asf$ users with a linear $t$-subset placement, over a field $\FF_{\qsf_0}$, achieve $(\Msf,\Rsf)$, with a possibly nonlinear delivery. Then $(\Msf,\Rsf)$ is also achievable in the hotplug model with $\Asf$ active users out of $\Ksf$, for every $\Ksf\geq\Asf$, over any extension field $\FF_\qsf$ of $\FF_{\qsf_0}$ with $\qsf\geq\Ksf-1$.
\end{thm}

\begin{cor}[Application of Theorem~\ref{thm:RM}: YMA in the hotplug model]
\label{cor:YMA}
For every $\Ksf\geq\Asf$, the lower convex envelope of the YMA points~\eqref{eq:YMA} of a classical system with $\Asf$ users is achievable in the hotplug model with $\Asf$ active users out of $\Ksf$.
\end{cor}
\begin{proof}[Proof of Corollary~\ref{cor:YMA}]
The point $(0,\rsf)$ is trivial. For $t\in[\Asf]$, apply Theorem~\ref{thm:RM} to the MAN placement with parameter $t$. Finally, use memory sharing, which is compatible with the hotplug model since the placements do not depend on the active users.
The subpacketization is the same as in the classical scheme.
\end{proof}

\begin{rem}[Relation to the Reed--Solomon embedding]
For $t=1$, Theorem~\ref{thm:RM} recovers~\cite[Theorem~2]{DT2027ICCsubmission} for row-wise placements with a {\it linear cache function}. The latter is more general: since for $t=1$ the restriction to the user's hyperplane is a single coded row, the cache may be any function of it, linear or not. For $t\geq2$, the restriction determines the classical subfiles only after the active users are known, so the cache must be computable from the restriction without knowing them; the linearity in Definition~\ref{def:tsubset} guarantees this. Hence the results of this note do not supersede~\cite[Theorem~2]{DT2027ICCsubmission}.
\end{rem}

\begin{rem}
\label{rem:consequences}
Corollary~\ref{cor:YMA}, together with~\cite[Corollary~1]{DT2027ICCsubmission}, shows that at every memory size $\Msf=\Nsf t/\Asf$, $t\in[\Asf-1]$, with $\min(\Asf,\Nsf)\geq2$ and $\Ksf>\Asf$, coded placement attains a load that no uncoded placement can attain in the hotplug model. For example, for $(\Asf,\Ksf,\Nsf)=(4,5,3)$ and $\Msf=\frac32$ (i.e., $t=2$), every uncoded placement has load at least $\frac34$ (memory sharing between the YMA+ points $(\frac65,1)$ and $(\frac95,\frac12)$~\cite{DT2027ICCsubmission}), whereas Corollary~\ref{cor:YMA} attains $\frac23$.
Theorem~\ref{thm:RM} also recovers the MDS-based schemes of~\cite{MT2022hotplug,MT2024arxiv} that attain YMA points, $t=1$ and $t=\Asf-1$ (HT3, see Section~\ref{sec:Reed--Muller t=1 HT3}), and, for $t\geq2$, it attains the YMA load of~\cite[Scheme~new1]{MT2022hotplug} with the memory size of the classical system, rather than one that depends on $\Ksf$. Combined with the Reed--Solomon embedding of GV~\cite[Theorem~2]{DT2027ICCsubmission}, the hotplug model attains the lower convex envelope of the GV and YMA points of a classical system with $\Asf$ users, for every $\Ksf\geq\Asf$.
\end{rem}

\subsection{Part~2: generic embedding}
In Definition~\ref{def:tsubset}, the same linear map acts on each subset separately. Some classical schemes with coded placement, such as the IE scheme, instead compress the whole MAN cache of a user jointly, mixing different files and different subsets.

\begin{defn}[Linear MAN-compression placement, generic correctness]
\label{def:compression}
For a classical system with $\Asf$ users, fix $t\in[\Asf]$, split each file as in Definition~\ref{def:tsubset}, and let
\begin{align}
\mathbf{s}_a:=\big(F_{n,\Wc}:\ n\in[\Nsf],\ \Wc\in\Omega^t_{[\Asf]},\ \Wc\ni a\big)\in\FF_\qsf^{1\times \Bsf\frac{\Nsf t}{\Asf}}
\end{align}
be the MAN cache of user $a$, i.e., the concatenation of its $\Nsf\binom{\Asf-1}{t-1}$ subfiles in a fixed order.
A placement is a \emph{linear MAN-compression placement} if
\begin{align}
Z_a=\mathcal{L}_a[\mathbf{s}_a],\quad a\in[\Asf],
\label{eq:compression}
\end{align}
where $\mathcal{L}_a:\FF_\qsf^{1\times \Bsf\frac{\Nsf t}{\Asf}}\to\FF_\qsf^{1\times \Bsf\Msf}$ is a linear operator, possibly different for each user.
A classical scheme with such a placement and with a linear delivery is \emph{generically correct with load $\Rsf$} if there are a field $\FF_{\qsf_0}$, a linear transmission $X_\dv=\mathcal{D}_\dv[F_1,\ldots,F_\Nsf]$ of $\Rsf\Bsf$ symbols with coefficients in $\FF_{\qsf_0}$ for every demand vector $\dv$, not depending on $(\mathcal{L}_1,\ldots,\mathcal{L}_\Asf)$, and a polynomial $g\not\equiv0$ with coefficients in $\FF_{\qsf_0}$ in the entries of the matrices of $\mathcal{L}_1,\ldots,\mathcal{L}_\Asf$, such that, over every extension field $\FF_\qsf$ of $\FF_{\qsf_0}$ and for all operators $\mathcal{L}_1,\ldots,\mathcal{L}_\Asf$ over $\FF_\qsf$ with $g(\mathcal{L}_1,\ldots,\mathcal{L}_\Asf)\neq0$, every user decodes its demanded file for every $\dv$.
\end{defn}

The MAN placement is the case in which $\mathcal{L}_a$ is the identity, i.e., $\Msf=\Nsf t/\Asf$, and the placement of Definition~\ref{def:tsubset} is the case in which $\mathcal{L}_a$ applies $\mathcal{L}$ to each subset $\Wc\ni a$ separately, the same for every user.

\begin{thm}[Generic Reed--Muller embedding]
\label{thm:generic}
Let a classical scheme for $\Asf$ users with a linear MAN-compression placement be generically correct with load $\Rsf$, with $\FF_{\qsf_0}$ and $g$ as in Definition~\ref{def:compression}. Then $(\Msf,\Rsf)$ is achievable in the hotplug model with $\Asf$ active users out of $\Ksf$, for every $\Ksf\geq\Asf$, over every extension field $\FF_\qsf$ of $\FF_{\qsf_0}$ with $\qsf\geq\Ksf-1$ and $\qsf>\binom{\Ksf}{\Asf}\deg(g)$. 
\end{thm}

\begin{cor}[Application of Theorem~\ref{thm:generic}: IE in the hotplug model]
\label{cor:IE}
Let $\Nsf\leq\Asf$. For every $\Ksf\geq\Asf$, the lower convex envelope of the IE points~\eqref{eq:IE} of a classical system with $\Asf$ users is achievable in the hotplug model with $\Asf$ active users out of $\Ksf$.
\end{cor}
\begin{proof}[Proof of Corollary~\ref{cor:IE}]
The point $(0,\Nsf)$, i.e., $t=0$, is trivial. For $t\in[\Asf]$, the IE placement is a linear MAN-compression placement with $\mathcal{L}_a$ the parity part of the systematic code that user $a$ applies to $\mathbf{s}_a$, and its delivery uses fixed MDS codes; by~\cite[Section~VI-D]{tian2018caching}, the scheme is generically correct. Apply Theorem~\ref{thm:generic}, and use memory sharing as in Corollary~\ref{cor:YMA}.
The subpacketization is the same as in the classical scheme.
\end{proof}

\subsection{Consequences}

\begin{cor}[Combination of Corollaries~\ref{cor:GV},~\ref{cor:YMA}, and~\ref{cor:IE}: the hotplug case of $\Asf=4$ active users and $\Nsf=2$ files]
\label{cor:42}
For every $\Ksf\geq\Asf=4$, $\Rsf^\star_{(4,\Ksf,2)}(\Msf)=\Rsf^\star_{(4,4,2)}(\Msf)$ for $\Msf\in[0,\frac14]\cup[\frac23,2]$, i.e., wherever the classical optimal tradeoff with $4$ users and $2$ files is known.
\end{cor}
\begin{proof}[Proof of Corollary~\ref{cor:42}]
On $[0,\frac14]$, the classical tradeoff is the line through $(0,2)$ and the CFL point $(\frac14,\frac32)$~\cite{chen2016fundamental}, achievable for every $\Ksf\geq4$ by Corollary~\ref{cor:GV}. On $[\frac23,2]$, its corner points are $(\frac23,1)$, attained by the IE scheme with $t=2$ and meeting the converse of~\cite{tian2018symmetry}, and the YMA points $(1,\frac23)$, $(\frac32,\frac14)$ and $(2,0)$~\cite{tian2018symmetry}; they are achievable for every $\Ksf\geq4$ by Corollaries~\ref{cor:IE} and~\ref{cor:YMA}.
\end{proof}

\section{Example $\Nsf=\Asf=3$}
\label{sec:example A=N=3}
Consider $\Nsf=3$ files, $A$, $B$ and $C$, and $\Asf=3$ active users out of $\Ksf\geq3$. Let $a_1,\ldots,a_\Ksf\in\FF_\qsf$ be distinct, which requires $\qsf\geq\Ksf$ (the point at infinity, discussed below, allows $\qsf\geq\Ksf-1$), and let the active users be $\Ic=\{i_1,i_2,i_3\}$, requesting $A$, $B$ and $C$, respectively. Indices of subfiles refer to the positions $j\in[3]$ of the active users in $\Ic$.

\subsection{$t=1$: Reed--Solomon embedding from~\cite{DT2027ICCsubmission}}
\label{sec:Reed--Solomon t=1}
The YMA point with $t=1$ is $(\Msf,\Rsf)=(1,1)$. Split each file into three rows, $A=(A_0,A_1,A_2)$, and so on. In the classical system, user $j\in[3]$ caches row $j-1$ of every file (we enumerated the subfile starting from $0$), a row-wise placement. In the hotplug model, user $k\in[\Ksf]$ caches
\begin{align}
Z_k=\big(p_A(a_k),\,p_B(a_k),\,p_C(a_k)\big),\qquad p_A(z):=A_0+A_1z+A_2z^2,
\label{eq:def p_A(z)}
\end{align}
and similarly for $p_B(z)$ and $p_C(z)$; each user caches one coded row per file, so that $\Msf=1$. The vector $(p_A(a_k) : k\in[\Ksf])$ is a codeword of a $[\Ksf,3]$ Reed--Solomon code. 
Once $\Ic=\{i_1,i_2,i_3\}$ is known, define the new library
\begin{align}
A'_j:=p_A(a_{i_j}),\quad B'_j:=p_B(a_{i_j}),\quad C'_j:=p_C(a_{i_j}),\qquad j\in[3].
\end{align}
The map $(A_0,A_1,A_2)\mapsto(A'_1,A'_2,A'_3)$ is given by a $3\times3$ Vandermonde matrix with distinct nodes, hence it is invertible. User $i_j$ caches exactly $(A'_j,B'_j,C'_j)$, i.e., the cache of classical user $j$ for the new library. The server sends the classical delivery for the new library
\begin{align}
A'_2+B'_1,\qquad A'_3+C'_1,\qquad B'_3+C'_2,
\end{align}
of load $\Rsf=1$. User $i_1$, for example, obtains $A'_2$ and $A'_3$ from the first two transmissions, recovers $(A'_1,A'_2,A'_3)$, and inverts the Vandermonde matrix to obtain $A$.

\subsection{$t=1$ revisited: the projective view}
\label{sec:Reed--Muller t=1}
The scheme in Section~\ref{sec:Reed--Solomon t=1} can be described geometrically, in a way that extends to $t\geq2$. Associate with user $k$ the normal vector $\hv_k:=(1,a_k)\in\FF_\qsf^2$ and the hyperplane $H_k:=\{\xv\in\FF_\qsf^2:\hv_k\cdot\xv=0\}$, which is the one-dimensional subspace spanned by $\pv_{\{k\}}:=(-a_k,1)$, a point of the projective line. Write file $A$ as a homogeneous polynomial of degree $\Asf-t=2$ in $t+1=2$ variables,
\begin{align}
f_A(x_0,x_1):=A_2x_0^2-A_1x_0x_1+A_0x_1^2;
\label{eq:def f_A(x0,x1)}
\end{align}
note that $f_A(\pv_{\{k\}})=p_A(a_k)$ with $p_A(z)$ defined in~\eqref{eq:def p_A(z)}. User $k$ caches the restriction of $f_A$ to $H_k$, i.e., its value at $\pv_{\{k\}}$, as the hyperplane $H_k$ is one-dimensional.
Once $\Ic=\{i_1,i_2,i_3\}$ is known, the subfiles of the new library are the values at the points $\pv_{\{j\}}$, $j\in\Ic$. The normal vector $\hv=(0,1)$, i.e., the point $\pv=(1,0)$ ``at infinity'', gives $f_A(1,0)=A_2$, and allows one more user, $\Ksf=\qsf+1$.

\subsection{$t=2$: Reed--Muller embedding recovering scheme HT3 in\cite{MT2024arxiv}}
\label{sec:Reed--Muller t=1 HT3}
The YMA point with $t=2$ is $(\Msf,\Rsf)=(2,\frac13)$. Associate with user $k$ the plane $H_k:=\{\xv\in\FF_\qsf^3:\hv_k\cdot\xv=0\}$ with normal vector $\hv_k:=(1,a_k,a_k^2)$. Any three of these normal vectors are linearly independent (Vandermonde), so any two planes meet in a line, spanned by
\begin{align}
\pv_{\{k,l\}}:=(a_ka_l,\,-(a_k+a_l),\,1),
\end{align}
the coefficient vector of $(z-a_k)(z-a_l)$; indeed $\hv_k\cdot\pv_{\{k,l\}}=a_ka_l-a_k(a_k+a_l)+a_k^2=0=\hv_l\cdot\pv_{\{k,l\}}$, while $\hv_j\cdot\pv_{\{k,l\}}=(a_j-a_k)(a_j-a_l)\neq0$ for $j\notin\{k,l\}$.

Split each file into $\binom32=3$ subfiles and write it as a homogeneous polynomial of degree $\Asf-t=1$ in $t+1=3$ variables, i.e., a linear form,
\begin{align}
f_A(\xv):=A_0x_0+A_1x_1+A_2x_2, \quad \xv=(x_0,x_1,x_2),
\label{eq:def f_A(x0,x1,x2)}
\end{align}
and similarly for $B$ and $C$. User $k$ caches the restriction of $f_A$ to its plane $H_k$, which is spanned by $\uv_{k,1}:=(-a_k,1,0)$ and $\uv_{k,2}:=(0,-a_k,1)$ (the coefficient vectors of $z-a_k$ and $z(z-a_k)$); that is, user $k$ caches
\begin{align}
f_A(\uv_{k,1})=A_1-a_kA_0,\qquad f_A(\uv_{k,2})=A_2-a_kA_1,
\end{align}
and similarly for $B$ and $C$, so that $\Msf=3\cdot\frac23=2$. This is the scheme HT3 of~\cite{MT2024arxiv}, in which user $k$ stores the projection of each file onto a plane.

Once $\Ic=\{i_1,i_2,i_3\}$ is known, the subfiles of the new library are the values at the intersection points of the active planes,
\begin{align}
A'_{\{j,l\}}:=f_A(\pv_{\{i_j,i_l\}}),\qquad \{j,l\}\in\Omega^2_{[3]}.
\end{align}
User $i_1$ knows $A'_{12}$ and $A'_{13}$: since $\pv_{\{k,l\}}=-a_l\uv_{k,1}+\uv_{k,2}$, it computes $f_A(\pv_{\{k,l\}})=-a_lf_A(\uv_{k,1})+f_A(\uv_{k,2})$ with $k=i_1$ (here everything is lienar). Conversely, these two values determine the restriction of $f_A$ to $H_{i_1}$, since $\pv_{\{i_1,i_2\}}$ and $\pv_{\{i_1,i_3\}}$ are independent vectors of $H_{i_1}$. Hence the cache of user $i_j$ is equivalent to the MAN cache of classical user $j$ with $t=2$ for the new library: both describe the restriction of the files to $H_{i_j}$, in two different bases. With $b_j:=a_{i_j}$, $j\in[3]$, the map $(A_0,A_1,A_2)\mapsto(A'_{12},A'_{13},A'_{23})$ has matrix
\begin{align}
\begin{bmatrix}b_1b_2&-(b_1+b_2)&1\\b_1b_3&-(b_1+b_3)&1\\b_2b_3&-(b_2+b_3)&1\end{bmatrix},
\qquad\text{with determinant}\quad -(b_1-b_2)(b_1-b_3)(b_2-b_3)\neq0,
\end{align}
so the new files are obtained from an invertible mapping. The server sends the classical delivery for the new library
\begin{align}
A'_{23}+B'_{13}+C'_{12},
\end{align}
of load $\Rsf=\frac13$. User $i_1$ knows $B'_{13}$ and $C'_{12}$, decodes $A'_{23}$, recovers $(A'_{12},A'_{13},A'_{23})$, and inverts the matrix above to obtain $A$.

\subsection{Conclusion for the case of three active users and three files}
For $t=0$ and $t=3$, the YMA points $(0,3)$ and $(3,0)$ are trivially achievable. Hence the YMA tradeoff of the classical system with $3$ users and $3$ files is achievable in the hotplug model for every $\Ksf\geq3$. While this example is easy to describe, it gives no new scheme compared to the state of the art~\cite{MT2024arxiv}.

\section{Example $\Nsf=3, \Asf=4$}
\label{sec:example N=3 A=4}
For the cases in Section~\ref{sec:example A=N=3}, the files are represented as linear forms. For $\Asf\geq4$ and $2\leq t\leq\Asf-2$, the files are homogeneous polynomials of degree $\Asf-t\geq2$ in $t+1\geq3$ variables, and the restriction of a file to a hyperplane is itself a polynomial of the same degree in $t$ variables (aka one less variable); these are the cases that were not covered by previous hotplug schemes at the memory size of the classical system. The smallest such case follows.

\subsection{Case $t=2$: a new point}
The YMA point with $\Asf=4$, $\Nsf=3$ and $t=2$ is $(\Msf,\Rsf)=(\frac32,\frac23)$. Associate with each user $k$ a distinct element $a_k$ of the field, which requires $\qsf\geq\Ksf$, and the normal vector $\hv_k:=(1,a_k,a_k^2)$; any three of these vectors are linearly independent (Vandermonde). The plane $H_k:=\{\xv\in\FF_\qsf^3:\hv_k\cdot\xv=0\}$ is spanned by $\uv_{k,1}:=(-a_k,1,0)$ and $\uv_{k,2}:=(0,-a_k,1)$, and two planes $H_k$ and $H_l$ meet in the line spanned by
\begin{align}
\pv_{\{k,l\}}:=(a_ka_l,\,-(a_k+a_l),\,1)=-a_l\uv_{k,1}+\uv_{k,2}=-a_k\uv_{l,1}+\uv_{l,2},
\end{align}
as in Section~\ref{sec:Reed--Muller t=1 HT3}.

Split each file into $\binom42=6$ subfiles and write it as a quadratic form in three variables, symbol by symbol,
\begin{align}
f_A(\xv):=A_{00}x_0^2+A_{01}x_0x_1+A_{02}x_0x_2+A_{11}x_1^2+A_{12}x_1x_2+A_{22}x_2^2,\quad \xv=(x_0,x_1,x_2),
\end{align}
and similarly for $B$ and $C$. User $k$ caches the restriction of $f_A$ to $H_k$, that is,
\begin{subequations}
\begin{align}
\tilde f_A(\lambda,\mu;a_k)
  &:=f_A(\lambda\uv_{k,1}+\mu\uv_{k,2}) \notag
\\&=\big(a_k^2A_{00}-a_kA_{01}+A_{11}\big)\lambda^2
\\&+\big(a_k^2A_{01}-a_k[A_{02}+2A_{11}]+A_{12}\big)\lambda\mu
\\&+\big(a_k^2A_{11}-a_kA_{12}+A_{22}\big)\mu^2,\qquad (\lambda,\mu)\in\FF_\qsf^2,
\end{align}
\label{eq:restrictionplane}
\end{subequations}
a binary quadratic form, i.e., a homogeneous polynomial of degree two in two variables: one variable fewer than $f_A$, and the same degree. Each coefficient in~\eqref{eq:restrictionplane} is a linear combination of the subfiles of $A$. Unlike the linear forms of Section~\ref{sec:Reed--Muller t=1 HT3}, the restriction $\tilde f_A(\lambda,\mu;a_k)$ is not determined by its values at the basis vectors $\uv_{k,1}$ and $\uv_{k,2}$; so, user $k$ stores its three coefficients in~\eqref{eq:restrictionplane}, and similarly for $B$ and $C$; with three subfiles per file the cache size is $\Msf=3\cdot\frac36=\frac32$.

Once the set $\Ic=\{i_1,i_2,i_3,i_4\}$ of four active users is known, the six subfiles of the new library are
\begin{align}
A'_{\{k,l\}}
  &:=f_A(\pv_{\{k,l\}})=f_A\big(a_ka_l,-(a_k+a_l),1\big)
\\&=\tilde f_A(-a_l,1;a_k)
\\&=\tilde f_A(-a_k,1;a_l),\qquad \{k,l\}\in\Omega^2_\Ic,
\end{align}
and similarly for $B'$ and $C'$. Since the planes $H_k$ are ``in general position,'' the map $A\mapsto A'$ is invertible by Lemma~\ref{lem:interp} below.
Hence user $k\in\Ic$ obtains the three subfiles $A'_{\{k,l\}}$, $l\in\Ic\setminus\{k\}$, by evaluating its cached form $\tilde f_A(\cdot,\cdot;a_k)$ at $(-a_l,1)$, and similarly for $B'$ and $C'$; by the equality $\tilde f_A(-a_l,1;a_k)=\tilde f_A(-a_k,1;a_l)$, users $k$ and $l$ obtain the same subfile $A'_{\{k,l\}}$, each from its own cache.
Hence the active users hold the MAN placement with $t=2$ for the new library, and the YMA delivery for four users and three files sends $\binom43-\binom{1}{3}=4$ messages of size $\frac16$, i.e., $\Rsf=\frac23$.
For example, for users $(i_1,i_2,i_3,i_4)=(1,2,3,4)$ with demand $(A,B,C,A)$, the server sends
\begin{subequations}
\begin{align}
X_{123}&=A'_{23}+B'_{13}+C'_{12}, \\ 
X_{124}&=A'_{24}+B'_{14}+A'_{12}, \\
X_{134}&=A'_{34}+C'_{14}+A'_{13}, \\
X_{234}&=B'_{34}+C'_{24}+A'_{23}.
\end{align}
\end{subequations}

\subsection{Comparison with other schemes}
For $\Asf=4$ active users, $\Nsf=3$ files and memory size $\Msf=\frac32$, Table~\ref{tab:comparison} compares the load of the Reed--Muller embedding with that of other hotplug schemes and with the cut-set bound, for $\Ksf=5$.

\begin{table}[h]
\centering
\caption{Loads at $\Msf=\frac32$ for $(\Asf,\Ksf,\Nsf)=(4,5,3)$.}
\label{tab:comparison}
\begin{tabular}{ll}
\hline
Scheme & Load at $\Msf=\frac32$, $\Ksf=5$\\\hline
Uncoded placement / YMA+~\cite[Theorem~1]{DT2027ICCsubmission} & $\geq\frac34$, (between $(\frac65,1)$ and $(\frac95,\frac12)$)\\
MDS on MAN placement~\cite[Scheme~new1]{MT2022hotplug} & $1$ (between $(\frac34,\frac32)$ and $(2,\frac23)$)\\
Reed--Muller embedding (Corollary~\ref{cor:YMA}) & $\frac23$\\
Cut-set bound & $\geq\frac12$\\\hline
\end{tabular}
\end{table}

The Reed--Muller embedding attains $\Rsf=\frac23$ for every $\Ksf\geq4$. In contrast, the two other schemes in Table~\ref{tab:comparison} degrade as $\Ksf$ grows: the optimal load under uncoded placement grows with $\Ksf$ toward the decentralized load $\frac78$; and the MDS layer of~\cite[Scheme~new1]{MT2022hotplug} attains the load $\frac23$ only at memory size $\Nsf\binom{\Ksf-1}{1}/\binom42=\frac{\Ksf-1}{2}$. Finally, the scheme HT3~\cite{MT2024arxiv} attains $(\frac94,\frac14)$, at a larger memory size, and the Reed--Solomon embedding of~\cite{DT2027ICCsubmission} applies only for $\Msf\leq\frac34$.

\section{Proof of Theorem~\ref{thm:RM}}
\label{sec:proof}
\subsection{Background on hyperplanes and their intersection points}
\begin{defn}[Hyperplanes in general position]
For $s\geq1$ and a nonzero $\hv\in\FF_\qsf^{s+1}$, let $H:=\{\xv\in\FF_\qsf^{s+1}:\hv\cdot\xv=0\}$, an $s$-dimensional subspace (a hyperplane of the projective space of dimension $s$) with normal vector $\hv$. A family of hyperplanes $\{H_k : k\in\Sc\}$ is said to be \emph{in general position} if any $\min(s+1,|\Sc|)$ of the normal vectors are linearly independent.
\end{defn}

Example of explicit construction: Let $a_1,\ldots,a_{\Ksf-1}\in\FF_\qsf$ be distinct, which requires $\qsf\geq\Ksf-1$, and, for $s\geq1$,
\begin{align}
\hv_k:=(1,a_k,\ldots,a_k^s),\ k\in[\Ksf-1],\qquad
\hv_\Ksf:=(0,\ldots,0,1).
\label{eq:normals}
\end{align}
For every $s\geq1$, the corresponding hyperplanes are in general position, i.e., any $\min(s+1,\Ksf)$ of these vectors are linearly independent: for $m$ vectors with $k<\Ksf$, their first $m$ coordinates form a nonsingular Vandermonde matrix, and if $\hv_\Ksf$ is included, the minor on the first $m-1$ coordinates and the last coordinate, expanded along $\hv_\Ksf$, leaves a Vandermonde determinant of order $m-1$. The field size does not depend on $s$. 

The following Lemma is an elementary linear algebra fact.
\begin{lem}[General position]
\label{lem:genpos}
Let $\{H_k\}_{k\in\Sc}$ be in general position in $\FF_\qsf^{s+1}$, with $|\Sc|\geq s$.
\begin{enumerate}[label=(\alph*)]
\item For $\Tc\in\Omega^s_\Sc$, $\cap_{k\in\Tc}H_k$ is one-dimensional, spanned by a vector $\pv_\Tc$; moreover, $\hv_\ell\cdot\pv_\Tc\neq0$ for $\ell\in\Sc\setminus\Tc$, and the points $\pv_\Tc$ are distinct.
\item If $s\geq2$ and $j\in\Sc$, then $\{H_k\cap H_j : k\in\Sc\setminus\{j\}\}$ are hyperplanes of $H_j$ in general position, and for $\Tc\in\Omega^{s-1}_{\Sc\setminus\{j\}}$ their intersection is spanned by $\pv_{\Tc\cup\{j\}}$.
\end{enumerate}
\end{lem}
\begin{proof}[Proof of Lemma~\ref{lem:genpos}]
(a) The $s$ vectors $\{\hv_k: k\in\Tc\}$ are linearly independent, so their common orthogonal complement is one-dimensional. If $\hv_k\cdot\pv_\Tc=0$ with $k\notin\Tc$, then the nonzero $\pv_\Tc$ would be orthogonal to $s+1$ independent vectors. If $\pv_\Tc$ and $\pv_{\Tc'}$ were proportional with $\Tc\neq\Tc'$, then $\pv_\Tc\in H_k$ for some $k\in\Tc'\setminus\Tc$, a contradiction.

(b) Linear forms on $H_j$ are restrictions of linear forms on $\FF_\qsf^{s+1}$, and two such forms have the same restriction if and only if their normal vectors differ by a multiple of $\hv_j$. Hence the restrictions of $\hv_k\cdot\xv$, $k\neq j$, are nonzero, and any $\min(s,|\Sc|-1)$ of them are independent if and only if the corresponding $\hv_k$, together with $\hv_j$, are independent, which holds. Finally, $\cap_{k\in\Tc}(H_k\cap H_j)=\cap_{k\in\Tc\cup\{j\}}H_k$.
\end{proof}

For $s\geq0$ and $\delta\geq0$, let $\Hom_{s+1,\delta}$ be the $\FF_\qsf$-vector space of homogeneous polynomials of degree $\delta$ in $s+1$ variables (including the zero polynomial). Its dimension is the number of monomials of degree $\delta$ in $s+1$ variables, that is,
\begin{align}
\dim\Hom_{s+1,\delta}=\binom{\delta+s}{s};
\label{eq:dim}
\end{align}
in particular, $\dim\Hom_{t+1,\Asf-t}=\binom{\Asf}{t}$ and $\dim\Hom_{t,\Asf-t}=\binom{\Asf-1}{t-1}$.

\begin{lem}[Interpolation at intersection points]
\label{lem:interp}
Let $\{H_k:k\in\Sc\}$ be in general position in $\FF_\qsf^{s+1}$, with $n:=|\Sc|\geq s\geq1$. Then the evaluation map
\begin{align}
\Hom_{s+1,n-s}\to\FF_\qsf^{\binom{n}{s}}:\ f\mapsto\big(f(\pv_\Tc):\Tc\in\Omega^s_\Sc\big)
\end{align}
is a linear bijection.
\end{lem}
\begin{proof}[Proof of Lemma~\ref{lem:interp}]
By~\eqref{eq:dim}, both spaces have dimension $\binom{n}{s}$, so it suffices to show that if $f(\pv_\Tc)=0$ for all $\Tc\in\Omega^s_\Sc$, then $f$ is the zero polynomial. 
Indeed, a linear map between spaces of equal finite dimension is bijective if and only if it is injective. It is injective if and only if its kernel is $\{0\}$, and the kernel consists of the $f$ with $f(\pv_\Tc)=0$ for all $\Tc$. So it is enough to show that only the zero polynomial vanishes at all the points.

We proceed by induction on $n$, proving the claim for all $s\in[n]$ at once. For $n=1$, necessarily $s=1=n$, which is covered by the first case below. Let $n\geq2$, assume the claim for $n-1$ hyperplanes and every $s\in[n-1]$ is true; let $s\in[n]$:

\emph{Case $s=n$:} $f$ is a constant, and it vanishes at the nonzero vector $\pv_\Sc$; hence $f=0$.

\emph{Case $s=1<n$:} $f$ is a binary form of degree $n-1$ with the $n$ pairwise non-proportional zeros $\pv_{\{k\}}$, $k\in\Sc$ (Lemma~\ref{lem:genpos}(a)); since a nonzero binary form of degree $n-1$ has at most $n-1$ zeros in the projective line, $f=0$.

\emph{Case $2\leq s\leq n-1$:} fix $j\in\Sc$ and change coordinates so that $\hv_j\cdot\xv$ is the first coordinate; write $f=(\hv_j\cdot\xv)\,g+\tilde{f}$, where $\tilde{f}$ does not depend on the first coordinate, so that $\tilde{f}$ is the restriction of $f$ to $H_j$, a homogeneous polynomial of degree $(n-1)-(s-1)$ in $s$ variables. By Lemma~\ref{lem:genpos}(b), the $n-1$ hyperplanes $H_k\cap H_j$, $k\in\Sc\setminus\{j\}$, of $H_j$ are in general position, and their intersection points are the points $\pv_{\Tc\cup\{j\}}$, $\Tc\in\Omega^{s-1}_{\Sc\setminus\{j\}}$, at which $\tilde{f}$ vanishes. By the induction hypothesis, applied to these $n-1$ hyperplanes with $s-1\in[n-1]$, $\tilde{f}=0$, so that $f=(\hv_j\cdot\xv)\,g$ with $g\in\Hom_{s+1,n-s-1}$. For $\Tc\in\Omega^s_{\Sc\setminus\{j\}}$, $\hv_j\cdot\pv_\Tc\neq0$ by Lemma~\ref{lem:genpos}(a), hence $g(\pv_\Tc)=0$. By the induction hypothesis, applied to the $n-1$ hyperplanes $\{H_k:k\in\Sc\setminus\{j\}\}$, which are in general position, with $s\in[n-1]$, $g=0$. Hence $f=0$.
\end{proof}


\subsection{The embedding and proof of Theorem~\ref{thm:RM}}
\begin{proof}[Proof of Theorem~\ref{thm:RM}]
\emph{Field.} Since the classical scheme is correct for every realization of the library, we may run $\mu$ instances of it in parallel and view $\mu$ symbols of $\FF_{\qsf_0}$ as one symbol of $\FF_\qsf$, $\qsf=\qsf_0^\mu\geq\Ksf-1$. The matrix of $\mathcal{L}$ in~\eqref{eq:tsubset} has entries in $\FF_{\qsf_0}$; applied to vectors over $\FF_\qsf$, it is $\FF_\qsf$-linear and acts on each of the $\mu$ instances as in the classical scheme.

\emph{Trivial case.} For $t=\Asf$, the only subset is $\Wc=[\Asf]$, so every user caches $\mathcal{L}[\fv_{[\Asf]}]$, which does not depend on the user, and the classical scheme applies unchanged. Let $t\in[\Asf-1]$, and let $\{H_k:k\in[\Ksf]\}$ be hyperplanes of $\FF_\qsf^{t+1}$ in general position, e.g., with the normal vectors~\eqref{eq:normals} with $s=t$.

\emph{Files as polynomials.} Index the $\binom{\Asf}{t}$ subfiles of each file by the monomials of degree $\Asf-t$ in $t+1$ variables, and let $f_n\in\Hom_{t+1,\Asf-t}$ be the homogeneous polynomial whose coefficients are the subfiles of $F_n$, so that $f_n(\xv)\in\FF_\qsf^{1\times\Bsf/\binom{\Asf}{t}}$ for every $\xv\in\FF_\qsf^{t+1}$. Let $\ell$ be the homogeneous polynomial of degree $\Asf-t$ whose coefficient of each monomial is $\mathcal{L}$ applied to the concatenation of the coefficients of that monomial in $f_1,\ldots,f_\Nsf$\footnote{For the MAN placement, $\mathcal{L}$ is the identity and $\ell(\xv)=[f_1(\xv),\ldots,f_\Nsf(\xv)]$, so that each user caches the restriction of every file separately, as in the examples of Sections~\ref{sec:example A=N=3} and~\ref{sec:example N=3 A=4}; for the CFL placement, $\ell(\xv)=f_1(\xv)+\cdots+f_\Nsf(\xv)$.}. Since evaluation combines coefficients linearly with scalar weights, and $\mathcal{L}$ is linear,
\begin{align}
\ell(\xv)=\mathcal{L}\big[f_1(\xv),\ldots,f_\Nsf(\xv)\big]\in\FF_\qsf^{1\times\Msf\Bsf/\binom{\Asf-1}{t-1}},\qquad \xv\in\FF_\qsf^{t+1}.
\label{eq:ell}
\end{align}

\emph{Placement.} Let $\{\uv_{k,1},\ldots,\uv_{k,t}\}$ be a basis of $H_k$; for the normal vectors~\eqref{eq:normals}, one may take $\uv_{k,j}$ as the coefficient vector of $z^{j-1}(z-a_k)$, $j\in[t]$, for $k\in[\Ksf-1]$, and, for $k=\Ksf$, the vector with a one in coordinate $j-1$ and zeros elsewhere (coordinates indexed by $[0:t]$), since $H_\Ksf=\{\xv:x_t=0\}$. User $k\in[\Ksf]$ caches the coefficients of
\begin{align}
\tilde\ell_k(\boldsymbol{\lambda}):=\ell\Big(\sum_{j\in[t]}\lambda_j\uv_{k,j}\Big),\qquad \boldsymbol{\lambda}\in\FF_\qsf^t,
\end{align}
a row vector of homogeneous polynomials of degree $\Asf-t$ in $t$ variables; by~\eqref{eq:dim}, these are $\binom{\Asf-1}{t-1}$ coefficients of $\Msf\Bsf/\binom{\Asf-1}{t-1}$ symbols each, i.e., $\Msf\Bsf$ symbols, as in Definition~\ref{def:tsubset}.

\emph{New library.} Given the set of active users $\Ic$, define
\begin{align}
F'_{n,\Wc}:=f_n(\pv_\Wc),\quad n\in[\Nsf],\ \Wc\in\Omega^t_\Ic,
\end{align}
and $F'_n:=(F'_{n,\Wc}:\Wc\in\Omega^t_\Ic)$ and $\fv'_\Wc:=[F'_{1,\Wc},\ldots,F'_{\Nsf,\Wc}]$.
By Lemma~\ref{lem:interp}, applied to the $\Asf$ hyperplanes $\{H_k:k\in\Ic\}$ with $s=t$, the map $F_n\mapsto F'_n$ is a bijection, the same for every file.

\emph{Caches of the active users.} For $k\in\Wc$, $\pv_\Wc\in H_k$; hence user $k\in\Ic$ obtains $\mathcal{L}[\fv'_\Wc]=\ell(\pv_\Wc)$ by evaluating its cached restriction, for every $\Wc\in\Omega^t_\Ic$ with $\Wc\ni k$. Thus the active users hold the placement of Definition~\ref{def:tsubset} for the library $(F'_n : n\in[\Nsf])$, with the users in $\Ic$ in place of the classical users $[\Asf]$, through any bijection between $\Ic$ and $[\Asf]$.

\emph{Delivery and decoding.} The server sends the classical transmission for the library $(F'_n : n\in[\Nsf])$ and the demands $\dv[\Ic]$, of load $\Rsf$. Since the classical scheme is correct for every realization of the library, every active user $k$ decodes $F'_{d_k}$ as the corresponding classical user, and recovers $F_{d_k}$ by inverting the bijection $F_n\mapsto F'_n$.
\end{proof}

\subsection{Equivalent View: Symmetric Polynomials}
\label{sec:sym}
This section gives an equivalent description of the code of Section~\ref{sec:proof} that does not use hyperplanes, together with the explicit formula for the intersection points used earlier; it is not needed for the proofs. We use the normal vectors $\hv_k=(1,a_k,\ldots,a_k^t)$ in~\eqref{eq:normals} with distinct $a_k$ for all $k\in[\Ksf]$, i.e., without the point at infinity, which requires $\qsf\geq\Ksf$.

A vector $\xv=(x_0,\ldots,x_t)$ lies in $H_k$ if and only if the polynomial $\sum_{i=0}^tx_iz^i$ vanishes at $z=a_k$. Hence, for $\Wc\in\Omega^t_{[\Ksf]}$, the intersection point $\pv_\Wc$ is the coefficient vector of $\prod_{k\in\Wc}(z-a_k)$, i.e.,
\begin{align}
\pv_\Wc=\Big((-1)^{t-i}\,e_{t-i}(a_\Wc)\Big)_{i=0}^{t},
\label{eq:pT}
\end{align}
where $e_r$ is the elementary symmetric polynomial (a symmetric polynomial is a polynomial in several variables that does not change when the variables are permuted) of degree $r$ in $t$ variables ($e_0=1$) and $a_\Wc:=(a_k:k\in\Wc)$, in any order. Similarly, $\uv_{k,j}$, the coefficient vector of $z^{j-1}(z-a_k)$, $j\in[t]$, is a basis of $H_k$, and for $\Wc\ni k$, $\pv_\Wc=\sum_{j\in[t]}\lambda_j\uv_{k,j}$, where $\boldsymbol\lambda$ is the coefficient vector of $\prod_{l\in\Wc\setminus\{k\}}(z-a_l)$.

Let $\Sym_{t,\delta}$ be the space of symmetric polynomials in $t$ variables with degree at most $\delta$ in each variable. For $f\in\Hom_{t+1,\Asf-t}$, let
\begin{align}
\sigma(f)(y_1,\ldots,y_t):=f\Big(\big((-1)^{t-i}e_{t-i}(y_1,\ldots,y_t)\big)_{i=0}^{t}\Big),
\end{align}
so that, by~\eqref{eq:pT}, $f(\pv_\Wc)=\sigma(f)(a_\Wc)$. The map $\sigma$ is an isomorphism from $\Hom_{t+1,\Asf-t}$ onto $\Sym_{t,\Asf-t}$: each $e_r$ has degree at most one in each variable, so $\sigma(f)$ has degree at most $\Asf-t$ in each variable; $\sigma$ is injective, because $e_1,\ldots,e_t$ are algebraically independent and $f$ is homogeneous, hence determined by its values with $x_t=e_0=1$; and both spaces have dimension $\binom{\Asf}{t}$, since a basis of $\Sym_{t,\delta}$ is given by the monomial symmetric polynomials indexed by $\delta\geq c_1\geq\cdots\geq c_t\geq0$, and there are $\binom{\delta+t}{t}$ of them.

Therefore, applying $\sigma$ to each component of $\ell$ in~\eqref{eq:ell}, the code can be described as follows: the files are vectors of symmetric polynomials in $t$ variables, with degree at most $\Asf-t$ in each variable; the subfile indexed by $\Wc$ is the evaluation at the set $a_\Wc$ of $t$ distinct field elements; and the cache of user $k$ is the polynomial obtained by fixing one variable to $a_k$, which lies in $\Sym_{t-1,\Asf-t}$. In this view, Lemma~\ref{lem:interp} states that a polynomial in $\Sym_{t,\Asf-t}$ is determined by its values at the $t$-subsets of any $\Asf$ distinct field elements. For $t=1$, $\Sym_{1,\Asf-1}$ is the space of univariate polynomials of degree at most $\Asf-1$, and the construction is the Reed--Solomon code of Section~\ref{sec:Reed--Solomon t=1}.

For the example of Section~\ref{sec:Reed--Muller t=1 HT3}, $t=2$ and $f_A(\xv)=A_0x_0+A_1x_1+A_2x_2$; then
\begin{align}
\sigma(f_A)(y_1,y_2)=A_0\,y_1y_2-A_1(y_1+y_2)+A_2,\qquad
\sigma(f_A)(a_k,y)=(A_2-a_kA_1)+(a_kA_0-A_1)\,y,
\end{align}
so that the cache of user $k$, $(A_1-a_kA_0,\,A_2-a_kA_1)$, is, up to sign and order, the pair of coefficients of the linear polynomial $\sigma(f_A)(a_k,y)$.
For the example of Section~\ref{sec:example N=3 A=4}, $t=2$, $\Asf=4$ and $f_A$ is a quadratic form; then
\begin{align}
\sigma(f_A)(y_1,y_2)
&=A_{00}\,y_1^2y_2^2-A_{01}\,y_1y_2(y_1+y_2)+A_{02}\,y_1y_2+A_{11}\,(y_1+y_2)^2-A_{12}\,(y_1+y_2)+A_{22},
\\
\sigma(f_A)(a_k,y)
&=\big(a_k^2A_{00}-a_kA_{01}+A_{11}\big)y^2-\big(a_k^2A_{01}-a_k[A_{02}+2A_{11}]+A_{12}\big)y+\big(a_k^2A_{11}-a_kA_{12}+A_{22}\big),
\end{align}
i.e., $\sigma(f_A)(a_k,y)=\tilde f_A(-y,1;a_k)$: the cache of user $k$ consists of the three coefficients of a univariate quadratic polynomial, and the subfile shared by users $k$ and $l$ is $A'_{\{k,l\}}=\sigma(f_A)(a_k,a_l)=\sigma(f_A)(a_l,a_k)$.

The symmetric polynomial view makes the Reed--Solomon case transparent; the projective view allows any hyperplanes in general position, includes the point at infinity (hence $\qsf\geq\Ksf-1$ rather than $\qsf\geq\Ksf$), and identifies the code as a projective Reed--Muller code~\cite{lachaud1990,sorensen1991} restricted to the intersection points of the users' hyperplanes.

\section{Exact versus Generic Embedding, and Proof of Theorem~\ref{thm:generic}}
\label{sec:generic}

\subsection{Why an exact embedding is limited to Definition~\ref{def:tsubset}}
Let $t\geq2$ (for $t=1$, see~\cite[Theorem~2]{DT2027ICCsubmission}). In the construction of Section~\ref{sec:proof}, write the restriction of the files to $H_k$ as the row vector $\mathbf{r}_k\in\FF_\qsf^{1\times\Nsf t\Bsf/\Asf}$ of its coefficients. For an active set $\Ic\ni k$, let $\mathbf{s}'_k$ be the MAN cache of user $k$ for the new library, i.e., the concatenation of the subfiles $F'_{n,\Wc}$, $n\in[\Nsf]$, $\Wc\in\Omega^t_\Ic$, $\Wc\ni k$. Then
\begin{align}
\mathbf{s}'_k=\beta_{\Ic,k}\big[\mathbf{r}_k\big],
\label{eq:reinterp}
\end{align}
where $\beta_{\Ic,k}$ evaluates the restriction at the points $\pv_\Wc\in H_k$, $\Wc\in\Omega^t_\Ic$, $\Wc\ni k$. By Lemma~\ref{lem:genpos}(b) and Lemma~\ref{lem:interp}, applied on $H_k$ to the $\Asf-1$ hyperplanes $H_l\cap H_k$, $l\in\Ic\setminus\{k\}$, with $s=t-1$, the map $\beta_{\Ic,k}$ is a linear bijection; it acts in the same way on every file and every symbol position, and it depends on the other active users. Note: for $t=1$, $H_k$ is spanned by $\pv_{\{k\}}$, the restriction is the value at $\pv_{\{k\}}$, and $\beta_{\Ic,k}$ is the identity.

Consider a classical scheme with a linear MAN-compression placement as in~\eqref{eq:compression}. Within this construction, i.e., when the cache of user $k$ is a linear function of $\mathbf{r}_k$, an exact embedding with the same memory requires each user $k$ to store, before $\Ic$ is known, data from which it can compute the cache $\mathcal{L}_a[\mathbf{s}'_k]=\mathcal{L}_a\circ\beta_{\Ic,k}[\mathbf{r}_k]$ of the classical user $a$ whose role it plays. Since both have size $\Msf\Bsf$, if $\mathcal{L}_a$ has full rank, the linear functions of $\mathbf{r}_k$ given by $\mathcal{L}_a\circ\beta_{\Ic,k}$ must span the same space for every active set $\Ic\ni k$ and every role $a$. Hence: (i) the span of the linear functions given by $\mathcal{L}_a$ must be invariant under the re-interpolation maps $\beta_{\Ic,k}\circ\beta_{\Ic',k}^{-1}$; and (ii) it must not depend on $a$. The placements of Definition~\ref{def:tsubset} satisfy both: $\mathcal{L}_a$ applies the same operator $\mathcal{L}$ to each subset, while the re-interpolation maps combine different subsets with scalar coefficients, in the same way for every file and every symbol position, so they commute with $\mathcal{L}$. User and file symmetry, which can be assumed without loss of generality, gives invariance only under permutations of the labels, not under re-interpolation. A compression that mixes subsets, as in~\cite{tian2018caching}, need not satisfy~(i), since the re-interpolation maps change how the subsets are combined. This is why Theorem~\ref{thm:generic} does not embed the classical scheme itself, but, for each active set, a classical scheme of the same family with different coefficients.

\subsection{Proof of Theorem~\ref{thm:generic}}

Since $\mathbf{s}_a$ has $\Nsf t\Bsf/\Asf$ symbols, the placement in Definition~\ref{def:compression} is a compression of the MAN placement only when $\Msf\leq\Nsf t/\Asf$; for $\Msf\geq\Nsf t/\Asf$ the user may as well store $\mathbf{s}_a$, which is the MAN placement.

The proof uses the following lemma, as in~\cite{tian2018caching}.
\begin{lem}[Schwartz--Zippel~\cite{schwartz1980,zippel1979}]
\label{lem:SZ}
Let $P\in\FF[x_1,\ldots,x_n]$ be a nonzero polynomial of total degree $d$ over a field $\FF$, and let $S\subseteq\FF$ be finite. If $x_1,\ldots,x_n$ are drawn independently and uniformly from $S$, then
\begin{align}
\Pr\big[P(x_1,\ldots,x_n)=0\big]\leq\frac{d}{|S|}.
\end{align}
In particular, if $|S|>d$, there exists $(x_1,\ldots,x_n)\in S^n$ with $P(x_1,\ldots,x_n)\neq0$.
\end{lem}

\begin{proof}[Proof of Theorem~\ref{thm:generic}]

\emph{Files as polynomials.} As in the proof of Theorem~\ref{thm:RM}, let $\{H_k:k\in[\Ksf]\}$ be hyperplanes of $\FF_\qsf^{t+1}$ in general position, which requires $\qsf\geq\Ksf-1$, write each file as a homogeneous polynomial of degree $\Asf-t$ in $t+1$ variables, and let $\mathbf{r}_k\in\FF_\qsf^{1\times\Nsf t\Bsf/\Asf}$ be the coefficients of the restriction of the files to $H_k$.

\emph{Placement.} Let user $k$ cache $\Psi_k[\mathbf{r}_k]$, where $\Psi_1,\ldots,\Psi_\Ksf:\FF_\qsf^{1\times\Nsf t\Bsf/\Asf}\to\FF_\qsf^{1\times\Msf\Bsf}$ are linear operators chosen below.

\emph{Induced classical placement.} Fix $\Ic$ and a bijection $\pi_\Ic:\Ic\to[\Asf]$, so that active user $k$ plays the role of classical user $\pi_\Ic(k)$. With the new library $F'_{n,\Wc}=f_n(\pv_\Wc)$, $n\in[\Nsf], \ \Wc\in\Omega^t_\Ic$, of the proof of Theorem~\ref{thm:RM}, and by~\eqref{eq:reinterp}, user $k$ caches
\begin{align}
\Psi_k[\mathbf{r}_k]=\mathcal{L}^{(\Ic)}_{\pi_\Ic(k)}\big[\mathbf{s}'_k\big],\qquad \mathcal{L}^{(\Ic)}_{\pi_\Ic(k)}:=\Psi_k\circ\beta_{\Ic,k}^{-1},
\end{align}
i.e., the active users hold the linear MAN-compression placement~\eqref{eq:compression} with operators $(\mathcal{L}^{(\Ic)}_a:a\in[\Asf])$ for the library $(F'_n : n\in[\Nsf])$.

\emph{Choice of the operators.} For each $\Ic$, the map $(\Psi_k)_{k\in\Ic}\mapsto(\mathcal{L}^{(\Ic)}_a)_{a\in[\Asf]}$ is an invertible linear change of variables, since distinct active users have distinct operators and each $\beta_{\Ic,k}$ is a bijection. Hence $g_\Ic:=g(\mathcal{L}^{(\Ic)}_1,\ldots,\mathcal{L}^{(\Ic)}_\Asf)$ is a nonzero polynomial of degree $\deg(g)$ in the entries of $\Psi_1,\ldots,\Psi_\Ksf$, and so is $\prod_{\Ic}g_\Ic$, of degree $\binom{\Ksf}{\Asf}\deg(g)$. By Lemma~\ref{lem:SZ} with $S=\FF_\qsf$, there exist $\Psi_1,\ldots,\Psi_\Ksf$ at which $\prod_{\Ic}g_\Ic\neq0$ whenever $\qsf>\binom{\Ksf}{\Asf}\deg(g)$.

\emph{Delivery and decoding.} Given $\Ic$ and $\dv[\Ic]$, the server sends the transmission $\mathcal{D}_{\dv'}[F'_1,\ldots,F'_\Nsf]$ of $\Rsf\Bsf$ symbols, where $\dv'$ is the classical demand vector in which classical user $\pi_\Ic(k)$ requests $d_k$. Since $g(\mathcal{L}^{(\Ic)}_1,\ldots,\mathcal{L}^{(\Ic)}_\Asf)\neq0$, the operators are over the extension $\FF_\qsf$ of $\FF_{\qsf_0}$, and the classical scheme is correct for every realization of the library, active user $k$ decodes $F'_{d_k}$ as classical user $\pi_\Ic(k)$, and recovers $F_{d_k}$ by inverting the bijection $F_n\mapsto F'_n$ of the proof of Theorem~\ref{thm:RM}.
\end{proof}

\subsection{Example: IE with $\Asf=4$ and $\Nsf=2$}
For $(\Asf,\Nsf)=(4,2)$ and $t=2$, \eqref{eq:IE} gives $(\Msf,\Rsf)=(\frac23,1)$, which by Corollary~\ref{cor:IE} is achievable in the hotplug model, and which was left open in~\cite{DT2027ICCsubmission}.

In the classical scheme, each file is split into $\Bsf=\binom42=6$ symbols indexed by $\Wc\in\Omega^2_{[4]}$, $\mathbf{s}_a\in\FF_\qsf^{1\times6}$ collects the three symbols of each file indexed by the subsets containing $a$, and user $a$ caches $\mathcal{L}_a[\mathbf{s}_a]\in\FF_\qsf^{1\times4}$.

In the hotplug model, the files are quadratic forms in three variables, as in Section~\ref{sec:example N=3 A=4}. User $k$ computes the restriction $\mathbf{r}_k\in\FF_\qsf^{1\times6}$ of both files to $H_k$, i.e., three coefficients per file, and caches $\Psi_k[\mathbf{r}_k]\in\FF_\qsf^{1\times4}$. Theorem~\ref{thm:RM} would store all of $\mathbf{r}_k$, at memory $\Msf=\Nsf t/\Asf=1$; here the user keeps a $4$-dimensional compression of it, and Theorem~\ref{thm:generic} shows that a generic compression is correct for every active set.


It was noted in~\cite{DT2027ICCsubmission} that this case hinges on embedding the IE scheme~\cite{tian2018caching}. Corollary~\ref{cor:42} settles it: the hotplug tradeoff with $4$ active users and $2$ files equals the classical one wherever the latter is known, for every $\Ksf\geq4$.

Thus, for every $\Ksf\geq\Asf$, the hotplug model attains the lower convex envelope of the GV points~\cite[Theorem~2]{DT2027ICCsubmission}, the YMA points (Corollary~\ref{cor:YMA}) and, for $\Nsf\leq\Asf$, the IE points (Corollary~\ref{cor:IE}) of a classical system with $\Asf$ users.

\section{Answer to an Open Question in~\cite{DT2027ICCsubmission}: the case $\Asf=\Nsf=3, \Msf=3/5$}
\label{sec:open}

Cao and Xu characterizes the optimal tradeoff for the classical $(\Asf,\Nsf)=(3,3)$ model under linear coding both in the placement ad delivery~\cite{caoxu2020}, whose corner points are at $\Msf\in\{\frac13,\frac12,\frac35,1,2\}$; all of them except the point $(\frac35,\frac32)$ are attained in the hotplug model for every $\Ksf\geq3$ by Corollary~\ref{cor:GV} and Corollary~\ref{cor:YMA}, while the IE points add nothing here (the point with $t=2$ in~\eqref{eq:IE} gives $(\frac43,1)$, which is dominated by the existing schemes). 

The scheme of~\cite{caoxu2020} that attains $(\frac35,\frac32)$ operates on $\qsf=2$ and splits each file into $10$ subfiles (one for all users, one for each pair, two for each user), and its placement relies on a cyclic orientation of the users; such a scheme does not appear to be generically correct according to Definition~\ref{def:compression},
so Theorem~\ref{thm:generic} does not appear to apply to it. A different placement found by Claude Opus 5.5 model, however, attains the same point $(\frac35,\frac32)$ and embeds explicitly in the spirit of RWP.

\begin{prop}[Point $(\Msf,\Rsf)=(\frac35,\frac32)$ for $\Asf=\Nsf=3$]
\label{prop:N=A=3,M=3/5}
For every $\Ksf\geq3$, the point $(\Msf,\Rsf)=(\frac35,\frac32)$ is achievable in the hotplug model with $3$ active users out of $\Ksf\geq3$ and $3$ files. Hence, under linear coding, the optimal tradeoff of this hotplug model equals that of the classical system with $3$ users and $3$ files, for every $\Ksf\geq3$. For $\Msf\in[0,1/2]\cup[1,3]$ optimality holds without any restriction on the type of coding by the converse of~\cite{tian2018symmetry}.
\end{prop}
\begin{proof}[Proof of Proposition~\ref{prop:N=A=3,M=3/5}]
Here we work with $\qsf=3$. 

{\it Classical model:}
We split each file $X\in\{A,B,C\}$ into nine private subfiles $X_{k,[Y]}$, $k\in[3]$, $Y\in\{A,B,C\}$, and one common subfile $X_c$, each of $\Bsf/10$ symbols. For each cyclic triple $(X,Y,W)\in\{(A,B,C),(B,C,A),(C,A,B)\}$, user $k\in[3]$ caches
\begin{align}
X_{k,[W]}+Y_{k,[Y]}+Y_c,\qquad X_{k,[X]}-X_c+Y_{k,[W]},
\label{eq:place35}
\end{align}
i.e., six symbols of $\Bsf/10$.
The placement is row-wise up to one common subfile per file. The placement is invariant under every permutation of the users, and under every permutation $\sigma$ of the files that also permutes the labels in brackets, provided that, for odd $\sigma$, each common subfile is also replaced by its negative, i.e., $X_c\mapsto-\sigma(X)_c$; for example, the transposition $A\leftrightarrow B$ maps $(Z_{k,1},\ldots,Z_{k,6})$ to $(Z_{k,2},Z_{k,1},Z_{k,6},Z_{k,5},Z_{k,4},Z_{k,3})$. Since $X_c\mapsto-X_c$ is an invertible re-encoding of the file, these are symmetries of the scheme.

Every demand vector is equivalent to $AAA$, $ABB$ or $ABC$. For demand $AAA$, the server sends $A$, for a total of $10$ transmissions. For demands $ABB$ and $ABC$, the deliveries has $15$ transmissions, and the decoding steps are given in Appendix~\ref{app:35}. Hence, $\Rsf=\frac{15}{10}=\frac32$. 

{\it Hotplug model:}
For the hotplug model, write  $F_n = (\Wm_n,\cv_n), n\in[3]$, where row $k\in[3]$ matrix $\Wm_n \in \FF_\qsf^{3\times{3 \Bsf/10}}$ collects the private subfiles of user $k\in[3]$ and $\cv_n\in \FF_\qsf^{1\times{\Bsf/10}}$ is the common subfile. The cache of user $k\in[3]$ is $\Phi(\ev_k\Wm_1,\ev_k\Wm_2,\ev_k\Wm_3,\cv_1,\cv_2,\cv_3)$, with $\Phi$ the map in~\eqref{eq:place35} and $\ev_k$ the standard vector with a $1$ in position $k$. As in~\cite[Theorem~2]{DT2027ICCsubmission}, user $k\in[\Ksf]$ of the hotplug model caches $\Phi(\gv_k\Wm_1,\gv_k\Wm_2,\gv_k\Wm_3,\cv_1,\cv_2,\cv_3)$, where $\gv_1,\ldots,\gv_\Ksf\in\FF_\qsf^{1\times3}$, $\qsf=3^\mu\geq\Ksf-1$, are such that any three are linearly independent. Given $\Ic$, the map $(\Wm_n,\cv_n)\mapsto(\Gm_\Ic\Wm_n,\cv_n)$, with $\Gm_\Ic$ the invertible matrix with rows $\gv_{i_1},\gv_{i_2},\gv_{i_3}$, is invertible and the same for every file; the cache of user $i_j$ is the cache of classical user $j$ for the new library, and the classical delivery, whose coefficients are in $\FF_3$, applies.
 
The lower bound under linear coding holds because a hotplug scheme restricted to any set of active users is a classical scheme. The converse of~\cite{caoxu2020} holds for linear schemes over every field.
\end{proof}

\begin{rem}
Note that for point $(\frac35,\frac32)$, $3\Rsf+4\Msf = 3\cdot\frac32+4\cdot\frac35=\frac{69}{10}<7$: by~\cite[Theorem~4]{DT2027ICCsubmission}, no row-wise placement can attain this point; however, a single common subfile per file is enough to overcome this limitation, while preserving the Reed--Solomon embedding for row-wise placement schemes.
\end{rem}

\section{Concluding Remarks}

Open questions are:
\begin{itemize}

\item
An explicit embedding of the IE schemes of~\cite{tian2018caching}, e.g., with a structured choice of the maps $\Psi_k$ such as a rank-metric code applied to the restriction, and a bound on the field size better than the union bound. 

\item
The embedding of schemes that are neither of the form of Definition~\ref{def:tsubset} nor generically correct as per Definition~\ref{def:compression}, and that generalize the placement~\eqref{eq:place35}, aka row-wise up to a common subfile, to other values of $\Nsf$ and $\Asf$ beyond $\Asf=\Nsf=3$. 

\item
Whether the optimal hotplug tradeoff depends only on $\Asf$ in general: every classical point known to us that is attained by a GV, YMA or IE scheme, and the corner points of the linear tradeoff for $\Asf=\Nsf=3$, are attained in the hotplug model for every $\Ksf\geq\Asf$, which supports a positive answer.

\end{itemize}

\appendix
\section{Deliveries for the placement in~\eqref{eq:place35}}
\label{app:35}
All operations are over $\FF_3$, which may be the key difference compared to the original scheme in\cite{caoxu2020}. 

\paragraph*{Placement:} Recall that the cache for user $k\in[3]$ is
\begin{subequations}
\begin{align}
Z_{k,1}&=A_{k,[C]}+B_{k,[B]}+B_c, \\
Z_{k,2}&=A_{k,[A]}-A_c+B_{k,[C]}, \\
Z_{k,3}&=B_{k,[A]}+C_{k,[C]}+C_c, \\
Z_{k,4}&=B_{k,[B]}-B_c+C_{k,[A]}, \\
Z_{k,5}&=C_{k,[B]}+A_{k,[A]}+A_c, \\
Z_{k,6}&=C_{k,[C]}-C_c+A_{k,[B]}.
\end{align}
\end{subequations}

\paragraph*{Delivery for demand $ABC$:} 
Users $1,2,3$ request $A,B,C$. The server sends
\begin{subequations}
\begin{align}
&X_1=A_{3,[B]},\\
&X_2=B_{1,[C]},\\
&X_3=C_{2,[A]},\\
&X_4=A_{2,[C]},\\
&X_5=B_{3,[A]},\\
&X_6=C_{1,[B]},\\
&X_7=A_{2,[A]}+A_{3,[A]},\\
&X_8=B_{3,[B]}+B_{1,[B]},\\
&X_9=C_{1,[C]}+C_{2,[C]},\\
&X_{10}=A_{2,[A]}-A_c,\\
&X_{11}=B_{3,[B]}-B_c,\\
&X_{12}=C_{1,[C]}-C_c,\\
&X_{13}=A_{2,[A]}+B_{3,[C]}-C_{2,[B]},\\
&X_{14}=B_{3,[B]}+C_{1,[A]}-A_{3,[C]},\\
&X_{15}=C_{1,[C]}+A_{2,[B]}-B_{1,[A]}.
\end{align}
\end{subequations}

\paragraph*{Decoding for demand $ABC$:} 
Each line of the decoding uses the transmissions $X_i$, the cache $Z_{k,j}$ of the user, and the subfiles decoded in the previous lines.
The joint cyclic shift of users, files and bracket labels ($k\to k+1$, $A\to B\to C\to A$) maps the demand $ABC$ to itself and permutes the transmissions within each row; hence it suffices to describe the decoding at user $1$.
Recall we work on $\qsf=3$, i.e., $\qsf x=0, \forall x\in\FF_{\qsf}$.
User $1$ does:
\begin{subequations}
\begin{align}
&A_{3,[B]}=X_1,\\
&A_{2,[C]}=X_4,\\
&A_{1,[B]}=-X_{12}+Z_{1,6},\\
&A_{1,[C]}=-X_8+X_{11}+Z_{1,1},\\
&A_{2,[B]}=X_{12}+X_{15}+Z_{1,3} = A_{2,[B]} + 3 C_{1,[C]},\\
&A_{1,[A]}=X_2+X_6-Z_{1,2}-Z_{1,5}= -2A_{1,[A]}= -2A_{1,[A]} +3 A_{1,[A]},\\
&A_c=X_2-Z_{1,2}+A_{1,[A]},\\
&A_{2,[A]}=X_{10}+A_c,\\
&A_{3,[A]}=X_7-A_{2,[A]},\\
&A_{3,[C]}=-X_8-X_{11}-X_{14}+Z_{1,4}=A_{3,[C]}-3 B_{3,[B]}.
\end{align}
\end{subequations}

\paragraph*{Delivery for demand $ABB$:} 
Users $1,2,3$ request $A,B,B$. The server sends
\begin{subequations}
\begin{align}
X_1&=B_{1,[A]}, \\
X_2&=B_{1,[C]},\\
X_3&=A_{1,[A]}+A_c, \\
X_4&=A_{1,[B]}+C_{1,[B]},\\
X_5&=A_{2,[C]}+A_{3,[C]}, \\
X_6&=B_{1,[B]}+B_c,\\
X_7&=A_{1,[A]}-A_{2,[A]}-A_{3,[A]}, \\
X_8&=A_{2,[B]}+A_{3,[B]}-C_c,\\
X_9&=B_{2,[A]}+B_{3,[A]}-B_{2,[B]}-B_{3,[B]}+B_{2,[C]}+B_{3,[C]},\\
X_{10}&=A_{2,[A]}-A_{3,[A]}, \\
X_{11}&=A_{2,[B]}-A_{3,[B]},\\
X_{12}&=A_{2,[C]}-A_{3,[C]}, \\
X_{13}&=B_{2,[A]}-B_{3,[A]},\\
X_{14}&=B_{2,[B]}-B_{3,[B]}, \\
X_{15}&=B_{2,[C]}-B_{3,[C]}.
\end{align}
Under the swap of users $2$ and $3$, $X_1,\ldots,X_9$ are invariant and $X_{10},\ldots,X_{15}$ change sign.
\end{subequations}

\paragraph*{Decoding for demand $ABB$: User $1$ (requests $A$):} 
\begin{subequations}
\begin{align}
A_{2,[C]}&=-X_5-X_{12}=-2A_{2,[C]}=A_{2,[C]}-3A_{2,[C]},\\
A_{3,[C]}&=-X_5+X_{12}=-2A_{3,[C]}=A_{3,[C]}-3A_{3,[C]},\\
A_{1,[C]}&=-X_6+Z_{1,1},\\
A_{1,[A]}&=X_2-X_3-Z_{1,2}=-2A_{1,[A]}=A_{1,[A]}-3A_{1,[A]},\\
A_c&=X_3-A_{1,[A]},\\
A_{1,[B]}&=X_3+X_4-Z_{1,5},\\
A_{2,[A]}&=X_7-X_{10}-A_{1,[A]}=-2A_{2,[A]}=A_{2,[A]}-3A_{2,[A]},\\
A_{3,[A]}&=-X_{10}+A_{2,[A]},\\
A_{2,[B]}&=-X_1-X_8-X_{11}+Z_{1,3}-Z_{1,6}+A_{1,[B]}=A_{2,[B]}-3A_{2,[B]}+3C_c,\\
A_{3,[B]}&=-X_{11}+A_{2,[B]}.
\end{align}
\end{subequations}

\paragraph*{Decoding for demand $ABB$: User $2$ (requests $B$):}
\begin{subequations}
\begin{align}
B_{1,[A]}&=X_1,\\
B_{1,[C]}&=X_2,\\
B_{2,[C]}&=X_3-X_7+X_{10}+Z_{2,2}=B_{2,[C]}+3A_{2,[A]},\\
B_{3,[C]}&=-X_{15}+B_{2,[C]},\\
B_{2,[A]}&=-X_8-X_{11}+Z_{2,3}-Z_{2,6}=B_{2,[A]}-3A_{2,[B]}+3C_c,\\
B_{3,[A]}&=-X_{13}+B_{2,[A]},\\
B_{2,[B]}&=X_9+X_{13}-X_{14}+X_{15}+B_{2,[C]}+B_{2,[A]}=B_{2,[B]}-3B_{2,[B]}+3B_{2,[A]}+3B_{2,[C]},\\
B_{3,[B]}&=-X_{14}+B_{2,[B]},\\
B_c&=X_5+X_{12}+Z_{2,1}-B_{2,[B]}=B_c+3A_{2,[C]},\\
B_{1,[B]}&=X_6-B_c.
\end{align}
\end{subequations}

\paragraph*{Decoding for demand $ABB$: User $3$ (requests $B$):} 
As user $2$, with the user indices $2$ and $3$ swapped and $X_{10},\ldots,X_{15}$ replaced by $-X_{10},\ldots,-X_{15}$.

\subsection{Cache structure at the point $(\frac35,\frac32)$}
\label{sec:structure35}

Lemma~\ref{lem:2R4M} next holds for every scheme, linear or not; by symmetry, the same bound holds with any user in place of user $1$ and any two distinct files in place of $F_1,F_2$.
\begin{lem}
\label{lem:2R4M}
For $(\Asf,\Nsf)=(3,3)$, every scheme with memory $\Msf$ and load $\Rsf$ satisfies
\begin{align}
4\Msf+2\Rsf\geq 
5+\max\left( \frac{1}{3}, \ \frac{H(Z_1|F_1,F_2)}{\Bsf} \right).
\label{eq:2R4M}
\end{align}
\end{lem}%

\begin{proof}
In~\cite{tian2018symmetry} it was proved that $6\Msf + 3 \Rsf \geq 8$; thus the first term in~\eqref{eq:2R4M}.

For the new term, the second term in~\eqref{eq:2R4M}, let $X:=X_{ABC}$, in which users $1,2,3$ request $A,B,C$, and $Y:=X_{ACB}$, in which users $1,2,3$ request $(F_1,F_2,F_3)=(A,C,B)$. From $(Z_1,Z_2,X)$, users $1$ and $2$ decode $A$ and $B$; from $(Z_1,Z_3,Y)$, users $1$ and $3$ decode $A$ and $B$. Hence
\begin{align}
H(Z_1)+H(Z_2)+H(X)&\geq H(Z_1,Z_2,X)=H(A,B,Z_1)+H(Z_2,X|A,B,Z_1),\\
H(Z_1)+H(Z_3)+H(Y)&\geq H(Z_1,Z_3,Y)=H(A,B,Z_1)+H(Z_3,Y|A,B,Z_1).
\end{align}
Summing these two inequalites and using
\begin{align}
H(Z_2,X|A,B,Z_1)+H(Z_3,Y|A,B,Z_1)\geq H(Z_2,Z_3,X,Y|A,B,Z_1), 
\end{align}
we obtain
\begin{align}
&2H(Z_1)+H(Z_2)+H(Z_3)+H(X)+H(Y)
\\&\geq H(A,B,Z_1)+H(A,B,Z_1,Z_2,Z_3,X,Y)
\\&= H(A,B,Z_1)+3\Bsf = H(Z_1|A,B)+5\Bsf,
\end{align}
since user $3$ decodes $C$ from $(Z_3,X)$, so that $(A,B,Z_1,Z_2,Z_3,X,Y)$ determines all three files. The claim follows from $H(Z_k)\leq\Msf\Bsf$, $H(X)\leq\Rsf\Bsf$ and $H(Y)\leq\Rsf\Bsf$.
\end{proof}

\begin{rem}[Cache structure at the point $(\frac35,\frac32)$ for $(\Asf,\Nsf)=(3,3)$]
\label{rem:structure35-1}
\begin{subequations}
In both the placement of~\cite{caoxu2020} and in~\eqref{eq:place35}, the caches are mutually independent and jointly independent of each single file, that is,
\begin{align}
H(Z_1) =H(Z_1|A)&=\tfrac{6}{10}\Bsf,\\
H(Z_1,Z_2) =H(Z_1,Z_2|A)&=\tfrac{12}{10}\Bsf,\\
H(Z_1,Z_2,Z_3) =H(Z_1,Z_2,Z_3|A)&=\tfrac{18}{10}\Bsf.
\end{align}
\label{eq:Cao-Xu and LP}
\end{subequations}
\begin{subequations}
Given two files, instead, the caches are dependent and jointly determine the third file, that is,
\begin{align}
H(Z_1|A,B)&=\tfrac{4}{10}\Bsf,\\
H(Z_1,Z_2|A,B)&=\tfrac{7}{10}\Bsf,\quad\text{i.e., } I(Z_1;Z_2|A,B)=\tfrac{1}{10}\Bsf,\\
H(Z_1,Z_2,Z_3|A,B)&=\Bsf=H(C).
\end{align}
\label{eq:Cao-Xu}
\end{subequations}
By Lemma~\ref{lem:2R4M}, at $\Msf=\frac35$ every scheme (linear or not) must satisfy $\Rsf\geq\frac{13}{10}+\frac{H(Z_1|A,B)}{2\Bsf}$. Hence every scheme with $H(Z_1|A,B)\geq\frac{4}{10}\Bsf$, such as Cao-Xu's scheme, must have $\Rsf\geq\frac32$; and every scheme with $\Rsf=\frac32-\epsilon$, $\epsilon>0$, must have $H(Z_k|F_i,F_j)\leq\big(\frac{4}{10}-2\epsilon\big)\Bsf$ for every user $k$ and every two files $F_i,F_j$; thus, each cache must store less about any file given the other two than in the known linear schemes.

The Linear Program (LP) that minimizes the load $\Rsf$ over Shannon inequalities only, with the six demand vectors that are permutations of $(A,B,C)$ and demand $(A,B,B)$, gives $\Rsf\geq\frac{22}{15}$ at $\Msf=\frac35$, which is consistent with Lemma~\ref{lem:2R4M} for $H(Z_1|A,B)=\frac13\Bsf$. An optimal solution of this ``7-demand-Shannon-only-LP'' gives the same entropies as in~\eqref{eq:Cao-Xu and LP} but conditioning in two files gives
\begin{subequations}
\begin{align}
H(Z_1|A,B)&=\tfrac13\Bsf,\\
H(Z_1,Z_2|A,B)&=\tfrac23\Bsf,\\
H(Z_1,Z_2,Z_3|A,B)&=\Bsf=H(C),
\end{align}
\label{eq:LP}
\end{subequations}
that is, given two files, the caches are mutually independent and split the third file into three disjoint parts.
Row-wise placement produces independent caches. Every row-wise placement (linear or not) satisfies $3\Rsf+4\Msf\geq7$~\cite{DT2027ICCsubmission}, i.e., $\Rsf\geq\frac{23}{15}>\frac{22}{15}$ at $\Msf=\frac35$. Hence, a scheme that attains the ``7-demand-Shannon-only-LP'' bound $\Rsf=\frac{22}{15}$ at $\Msf=\frac35$, if it exists, has a placement that is neither linear nor row-wise; thus, if it exists, the split of the third file among the caches must depend on the values of the two given files.

Table~\ref{tab:LPentropies} reports the conditional entropies of sets of caches and transmissions given no file, one file and two files.
The formnat is $(a;b;c)$, where $a$ is for an optimal solution of the LP with Shannon inequalities only; $b$ for an optimal solution of the LP with, in addition, the common information variable $K_1$ of~\cite{caoxu2020}, which is a function of $(Z_1,X_{BAC})$ and of $A$, which gives $\Rsf\geq\frac32$; and $c$ for the scheme in this Appendix which is equivalent to~\cite{caoxu2020}. The two LP were solved with the HiGHS solver, with the six permutations of $(A,B,C)$ and with $(A,B,B)$ as demand vectors, and with the symmetries of the problem that map these demand vectors and the variable $K_1$ to themselves. The optimal solutions of the programs are not unique, and Table~\ref{tab:LPentropies} reports the ones returned by the solver. 
At every optimal solution of the ``7-demand-Shannon-only-LP'', we have: every cache has $H(Z_k)=H(Z_k|F_i)=\Msf\Bsf$; every two caches are independent, unconditionally and given one file; the joint entropy of the three caches, unconditionally or given one file, is in a range $H(Z_1,Z_2,Z_3)/\Bsf, H(Z_1,Z_2,Z_3|A)/\Bsf \in [46/30,18/10]$, which is noted in Table~\ref{tab:LPentropies} as $18^*$;
and, given two files, the three caches are mutually independent.
\end{rem}

\begin{rem}
\label{rem:structure35-2}
A stronger form of the second bound in Lemma~\ref{lem:2R4M} is: with $\sum_{j\in[3]} H(Z_j) \leq 3\Msf\Bsf$ we have
\begin{align}
5 
\leq 2\Rsf + 3\Msf +\min_{k,\{i,j\}}\frac{I(Z_k;F_i,F_j)}{\Bsf}.
\end{align}
Since the files are independent, $I(Z_1;F_1,F_2)=I(Z_1;F_1)+I(Z_1;F_2)+I(F_1;F_2|Z_1)$; in the Cao-Xu's schemes and at every optimal solution of ``7-demand-Shannon-only-LP'', the caches are independent of each single file, so $I(Z_1;F_1,F_2)=I(F_1;F_2|Z_1)$; hence, Lemma~\ref{lem:2R4M} states that, to lower the load, every cache must create more dependence between every two files.
\end{rem}

\begin{table}
\centering
\caption{\small Here $\Bsf=10=H(F_n)$; we show $H(\text{row}\,|\,\text{column})$ for $(\Nsf,\Ksf)=(3,3)$ and $\Msf=\frac35\Bsf=6$; entries are $(a;b;c)$, with $a$ at an optimal point of the LP with Shannon inequalities only ($\Rsf\geq\frac{22}{15} = \frac{44}{3 \Bsf}$), $b$ at an optimal point of the LP with Shannon inequalities and the common information $K_1$ of~\cite{caoxu2020} ($\Rsf\geq\frac32 = \frac{15}{\Bsf}$), and $c$ for the achievable scheme of~\cite{caoxu2020}.}
\label{tab:LPentropies}
\begin{tabular}{l|ccc}
\hline
 & $\emptyset$ & $F_1$ & $F_1,F_2$ \\ \hline
\hline
$Z_1$                         & $(6;6;6)$                     & $(6;6;6)$                     & $(\frac{10}{3};4;4)$ \\
$Z_1,Z_2$                     & $(12;12;12)$                  & $(12;12;12)$                  & $(\frac{20}{3};7;7)$ \\
$Z_1,Z_2,Z_3$                 & $(18^*;18;18)$                & $(18^*;16;18)$                & $(10;10;10)$ \\ \hline
\hline
$X_{ABC}$                     & $(\frac{44}{3};15;15)$        & $(\frac{32}{3};11;11)$        & $(\frac{20}{3};7;7)$ \\
$Z_1,X_{ABC}$                 & $(\frac{62}{3};21;21)$        & $(\frac{32}{3};11;11)$        & $(\frac{20}{3};7;7)$ \\
$Z_1,Z_2,X_{ABC}$             & $(\frac{80}{3};27;27)$        & $(\frac{50}{3};17;17)$        & $(\frac{20}{3};7;7)$ \\
$Z_1,Z_2,Z_3,X_{ABC}$         & $(30;30;30)$                  & $(20;20;20)$                  & $(10;10;10)$ \\ \hline
\hline
$X_{ABB}$                     & $(\frac{44}{3};15;15)$        & $(8;8;8)$                     & $(\frac{2}{3};1;1)$ \\
$Z_1,X_{ABB}$                 & $(\frac{62}{3};21;21)$        & $(\frac{32}{3};11;11)$        & $(\frac{10}{3};4;4)$ \\
$Z_1,Z_2,X_{ABB}$             & $(\frac{80}{3};27;27)$        & $(\frac{50}{3};17;17)$        & $(\frac{20}{3};7;7)$ \\
$Z_1,Z_2,Z_3,X_{ABB}$         & $(30;30;30)$                  & $(20;20;20)$                  & $(10;10;10)$ \\ \hline
\hline
$X_{ABB},X_{ABC}$             & $(\frac{64}{3};21;23)$        & $(14;14;16)$                  & $(\frac{20}{3};7;7)$ \\
$Z_1,X_{ABB},X_{ABC}$         & $(24;24;26)$                  & $(14;14;16)$                  & $(\frac{20}{3};7;7)$ \\
$Z_1,Z_2,X_{ABB},X_{ABC}$     & $(\frac{80}{3};27;27)$        & $(\frac{50}{3};17;17)$        & $(\frac{20}{3};7;7)$ \\
$Z_1,Z_2,Z_3,X_{ABB},X_{ABC}$ & $(30;30;30)$                  & $(20;20;20)$                  & $(10;10;10)$ \\ \hline
\end{tabular}
\end{table}


\begin{thebibliography}{99}
\bibitem{MAN} M. A. Maddah-Ali and U. Niesen, ``Fundamental limits of caching,'' \emph{IEEE Trans. Inf. Theory}, vol.~60, no.~5, pp.~2856--2867, 2014.

\bibitem{YMA} Q. Yu, M. A. Maddah-Ali, and A. S. Avestimehr, ``The exact rate-memory tradeoff for caching with uncoded prefetching,'' \emph{IEEE Trans. Inf. Theory}, vol.~64, no.~2, pp.~1281--1296, 2018.

\bibitem{YMAfactor2} Q. Yu, M. A. Maddah-Ali, and A. S. Avestimehr, ``Characterizing the rate-memory tradeoff in cache networks within a factor of 2,'' \emph{IEEE Trans. Inf. Theory}, vol.~65, no.~1, pp.~647--663, 2019.

\bibitem{WTP} K. Wan, D. Tuninetti, and P. Piantanida, ``An index coding approach to caching with uncoded cache placement,'' \emph{IEEE Trans. Inf. Theory}, vol.~66, no.~3, pp.~1318--1332, 2020.

\bibitem{chen2016fundamental} Z. Chen, P. Fan, and K. B. Letaief, ``Fundamental limits of caching: Improved bounds for users with small buffers,'' \emph{IET Commun.}, vol.~10, no.~17, pp.~2315--2318, 2016.

\bibitem{gomez2018fundamental} J. G\'omez-Vilardeb\'o, ``Fundamental limits of caching: Improved rate-memory tradeoff with coded prefetching,'' \emph{IEEE Trans. Commun.}, vol.~66, no.~10, pp.~4488--4497, 2018.

\bibitem{tian2018caching} C. Tian and J. Chen, ``Caching and delivery via interference elimination,'' \emph{IEEE Trans. Inf. Theory}, vol.~64, no.~3, pp.~1548--1560, 2018.

\bibitem{tian2018symmetry} C. Tian, ``Symmetry, outer bounds, and code constructions: A computer-aided investigation on the fundamental limits of caching,'' \emph{Entropy}, vol.~20, no.~8, p.~603, 2018.

\bibitem{caoxu2020} D. Cao and Y. Xu, ``Characterizing linear memory-rate tradeoff of coded caching: The $(N,K)=(3,3)$ case,'' arXiv:2005.05548, 2020.

\bibitem{MT2022hotplug} Y. Ma and D. Tuninetti, ``On coded caching systems with offline users,'' in \emph{Proc. IEEE ISIT}, pp.~1133--1138, 2022; arXiv:2202.01299. 

\bibitem{MT2024arxiv} Y. Ma and D. Tuninetti, ``On coded caching systems with offline users, with and without demand privacy against colluding users,'' arXiv:2401.06894, 2024.

\bibitem{DT2027ICCsubmission} D. Tuninetti, ``Hotplug coded caching: Uncoded versus row-wise coded placements,'' submitted to ICC 2027 and available on arXiv:2611.03045, 2026. 

\bibitem{schwartz1980} J.~T. Schwartz, ``Fast probabilistic algorithms for verification of polynomial identities,'' \emph{J. ACM}, vol.~27, no.~4, pp.~701--717, 1980.

\bibitem{zippel1979} R.~Zippel, ``Probabilistic algorithms for sparse polynomials,'' in \emph{Proc. EUROSAM}, LNCS vol.~72, pp.~216--226, 1979.

\bibitem{kumar2021exact} 
Vijith Kumar K P, B.~K.~Rai, and T.~Jacob ``The exact rate memory tradeoff for small caches with coded placement,'' arXiv:2102.04797, 2021.


\bibitem{lachaud1990} G.~Lachaud, ``The parameters of projective Reed--Muller codes,'' \emph{Discrete Math.}, vol.~81, no.~2, pp.~217--221, 1990. 

\bibitem{sorensen1991} A.~B. S{\o}rensen, ``Projective Reed--Muller codes,'' \emph{IEEE Trans. Inf. Theory}, vol.~37, no.~6, pp.~1567--1576, 1991. 

\end{thebibliography}
\end{document}